\documentclass{article}

\usepackage[a4paper, margin=0.6in]{geometry}
\usepackage[utf8]{inputenc}
\usepackage[dvipsnames,table,xcdraw]{xcolor}
\usepackage[affil-it]{authblk}
\usepackage[misc]{ifsym}
\usepackage{multirow}
\renewcommand{\arraystretch}{1.5}
\usepackage{appendix,
verbatim,
listings,
braket,
enumitem, 
kantlipsum,
graphicx,
multicol,
mathtools,
csquotes,
amsmath,
mathtools,
url,
subfiles,
amsthm,
amssymb,
fdsymbol,
rotating,
tikz,
tcolorbox,
forest,
caption,
hyperref,
fontawesome,
xcolor,
sectsty,
arydshln,
ulem,
subcaption,
caption,
longtable,
qcircuit}

\tcbuselibrary{raster, skins, breakable, listings}
\def\titlefont{\color{RoyalPurple}}
\sectionfont{\color{RoyalPurple}}
\subsectionfont{\color{RoyalPurple}}
\subsubsectionfont{\color{RoyalPurple}}

\let\OLDthebibliography\thebibliography
\renewcommand\thebibliography[1]{
  \OLDthebibliography{#1}
  \setlength{\parskip}{0pt}
}

\title{\titlefont \textbf{\Large EQISA: Energy-efficient Quantum Instruction Set Architecture\\ using Sparse Dictionary Learning}}

\author[1,2]{Sibasish Mishra}
\author[1,2,3]{Aritra Sarkar}
\author[1,2]{Sebastian Feld}

\affil[1]{Quantum Computing Division, QuTech, The Netherlands}
\affil[2]{Department of Quantum \& Computer Engineering, Delft University of Technology, The Netherlands}
\affil[3]{Quantum Intelligence Alliance, Kolkata, India}

\affil[ \Letter ]{s.feld@tudelft.nl}

\date{}

\begin{document}


\maketitle

\begin{abstract}
The scalability of quantum computing in supporting sophisticated algorithms critically depends not only on qubit quality and error handling, but also on the efficiency of classical control, constrained by the cryogenic control bandwidth and energy budget. 
In this work, we address this challenge by investigating the algorithmic complexity of quantum circuits at the instruction set architecture (ISA) level.
We introduce an energy-efficient quantum instruction set architecture (EQISA) that synthesizes quantum circuits in a discrete Solovay-Kitaev basis of fixed depth and encodes instruction streams using a sparse dictionary learned from decomposing a set of Haar-random unitaries, followed by entropy-optimal Huffman coding and an additional lossless bzip2 compression stage. 
This approach is evaluated on benchmark quantum circuits demonstrating over 60\% compression of quantum instruction streams across system sizes, enabling proportional reductions in classical control energy and communication overhead without loss of computational fidelity. 
Beyond compression, EQISA facilitates the discovery of higher-level composable abstractions in quantum circuits and provides estimates of quantum algorithmic complexity. 
These findings position EQISA as an impactful direction for improving the energy efficiency and scalability of quantum control architectures.
\end{abstract}

\section{Introduction}
\label{sec_1}

The quantum computing~(QC) paradigm leverages principles from quantum mechanics to enable information processing~\cite{benioff1980computer,deutsch1985quantum} orchestrated via quantum algorithms.
This paradigm holds the potential to significantly reduce computational resource requirements~\cite{bernstein1993quantum} for specific problem domains~\cite{feynman1982simulating,shor2022early} compared to classical computation, which has led to burgeoning interest from both academic and industrial stakeholders.
QC is implemented physically by fabricating quantum processing units (QPU). 
Various technologies, such as superconducting circuits, trapped ions, photonics, and electron spins, are being developed in tandem towards a scalable, high-quality QPU~\cite{leymann2020bitter,ezratty2023we,waintal2024quantum} capable of embodying the well-studied theoretical advantage.

The present trajectory of quantum computing research can be broadly mapped onto four distinct avenues: (i) algorithms orchestrated via quantum circuits, both via human design \cite{shor2022early} and program synthesis \cite{sarkar2024automated}, are enabling the proliferation of QC across novel application domains, (ii) circuit-level compilation \cite{sarkar2024yaqq,steinberg2024lightcone} and pulse-level control \cite{fauquenot2024eo} emphasizes the system integration of software and hardware, (iii) quantum error handling~\cite{cao2021nisq} via detection, correction or mitigation focuses on improving the fidelity of logical operations, facilitating the realization of practical quantum advantage, and (iv) exploring new materials and designs for the fabrication of qubits~\cite{chae2024elementary} targeted at robust quantum information storage and manipulation while being scalable and controllable.
This project targets the circuit-level compilation avenue, specifically the instruction set architecture (ISA), the abstract model that defines the programmable interface of a computer.
In this work, we propose enhancing the efficiency of the quantum instruction set architecture (QISA) design and optimal control.
The community is taking initiatives to adopt standardized representations, such as OpenQASM3~\cite{openqasm3}, and to explore multi-level intermediate representations (MLIR) to streamline quantum programming and compilation~\cite{mlir_q}. 
Scalable and integrated cryogenic control architectures~\cite{cryo-cmos}, more efficient QISA by balancing resource trade-offs~\cite{anastasia}, and optimal control pulse shaping are among promising allied approaches. 

The QC stack's abstraction layers~\cite{bertels2020quantum,bertels2021quantum} use various representations for the operations, including a computationally universal set of gates, initialization, and measurements. 
Embracing concepts from resource theory and descriptive complexity, and drawing on the efficiency gains observed in code compression in embedded systems~\cite{waterman2011}, this endeavor seeks to harness similar principles for the circuit-level representation of quantum computation.
This research's central idea is to synthesize quantum circuits into a discrete basis and to find a compressed representation of the quantum instruction stream. 
With the growing need for higher computational power and the functional limits of conventional circuitry, the size of the instruction stream sent to the processor increases significantly, imposing substantial, unproductive overhead on system energy and processor core power. 
This problem is also ominously present in quantum systems. 
The growing number of qubits and the development of more sophisticated quantum algorithms and protocols, such as error-correcting schemes, intensify the need for effective control processes. 
Against this backdrop, we formulate the research question $\mathcal{RQ}$ of this study. 
The answer to this question is addressed from both theoretical and pragmatic perspectives, leading to two consequential contributions, $\mathcal{C}_1$ and $\mathcal{C}_2$, of this work.
\begin{enumerate}[leftmargin=1.2cm]
    \item[$\mathcal{RQ}$:] How can we compress the representation of decomposed quantum circuits?
    \item[$\mapsto$] How can we estimate the algorithmic complexity for quantum computation?
\begin{enumerate}[nolistsep,leftmargin=1.0cm]
    \item[$\mathcal{C}_1$:] Discovery of high-level quantum programming abstractions.
\end{enumerate}
    \item[$\mapsto$] How can we design a compressed quantum instruction set architecture?
\begin{enumerate}[nolistsep,leftmargin=1.0cm]
    \item[$\mathcal{C}_2$:] Optimization of the energy efficiency of quantum control.
\end{enumerate}
\end{enumerate}

\begin{figure}[hbt]
    \centering
    \includegraphics[width=0.7\textwidth, clip, trim={13cm 0cm 0cm 6cm}]{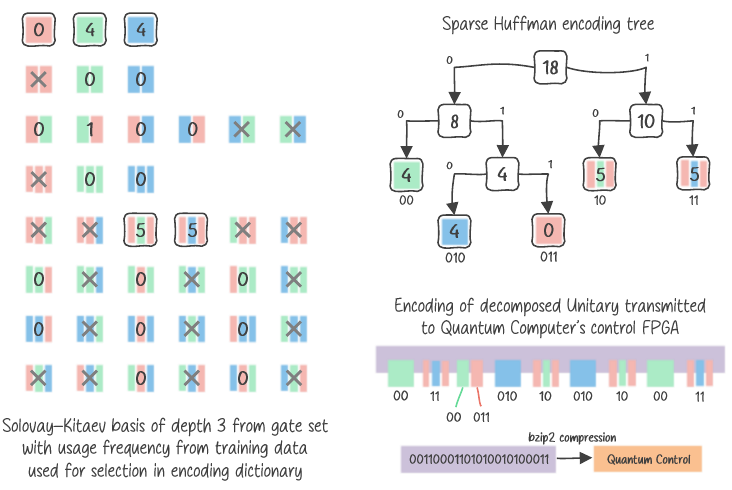}
    \caption{Overview of the proposed EQISA. The workflow is as follows: (i) A universal gate set (say, [$H$, $T$, $T^\dagger$], depicted in red, green, blue) is used to generate the Solovay-Kitaev basis of composite gates up to a specific depth (say, 3) based on required decomposition accuracy. (ii) Equivalent elements and identities are pruned (denoted by $\times$). (iii) The usage frequency for the remaining elements are determined by decomposing a set of Haar-random unitaries. (iv) High-frequency elements are used to construct a Huffman code. The original gate set is always included to decompose elements not included. (v) The EQISA thus constructed is deployed in the quantum compilation and control pipeline. The unitary of a quantum algorithm is decomposed into the basis elements. Elements not in the sparse dictionary are encoded in the original gate set. The bit sequence encodes the unitary and can be further compressed using bzip2. (vi) This stream is sent to the cryo-FPGA that decompresses and decodes the EQISA and dispatches the corresponding control signals for the quantum processor.}
    \label{fig:eqisa_overview}
\end{figure}

The proposed energy-efficient quantum instruction set architecture (EQISA) uses the Solovay-Kitaev basis of a predefined depth in the Solovay-Kitaev single-qubit decomposition algorithm, infers a sparse dictionary Huffman encoding from Haar-random training data, adds an additional bzip2 lossless compression pass, and evaluates the performance of pragmatic algorithms on benchmark quantum circuits.
The workflow of the EQISA is explained in Figure~\ref{fig:eqisa_overview}.
EQISA is demonstrated across three use cases: 
\begin{itemize}[nolistsep,noitemsep]
    \item[-] the optimization of the energy budget in quantum cryogenic control, 
    \item[-] the discovery of composable abstractions in quantum circuits, and 
    \item[-] the estimation of quantum algorithmic complexity. 
\end{itemize}


The rest of the article is organized as follows.
Section~\ref{resources} presents a survey of classical and quantum computational resources and metrics for estimating them individually or jointly. Thereafter, we justify our choice of a unitary matrix as the representation for describing quantum computation and provide an overview of methods for estimating description complexity based on algorithmic information theory, compression algorithms, and coding techniques.
Section~\ref{isa} presents the quantum stack in view of the instruction set architecture as our target layer to optimize.
We present the background on related techniques for instruction-set compression in classical computing.
In Section~\ref{ud}, multi-qubit and single-qubit unitary decomposition algorithms are presented, focusing on the Solovay-Kitaev decomposition as the guiding structure for our proposal.
Section~\ref{method} presents the proposed encoding methods and benchmarking results against random and algorithmic datasets. Thereafter, the introduced sparse dictionary learning variant of EQISA is augmented with the bzip2 lossless compression.
In Section~\ref{applications}, three applications of the proposed EQISA are presented.
Section~\ref{conclusion} concludes the article with suggestions for future work.

\section{Computational resources}
\label{resources}

\begin{figure}[htb]
    \centering
    \includegraphics[width=\textwidth,trim={3.4cm 0cm 0 7.7cm},clip]{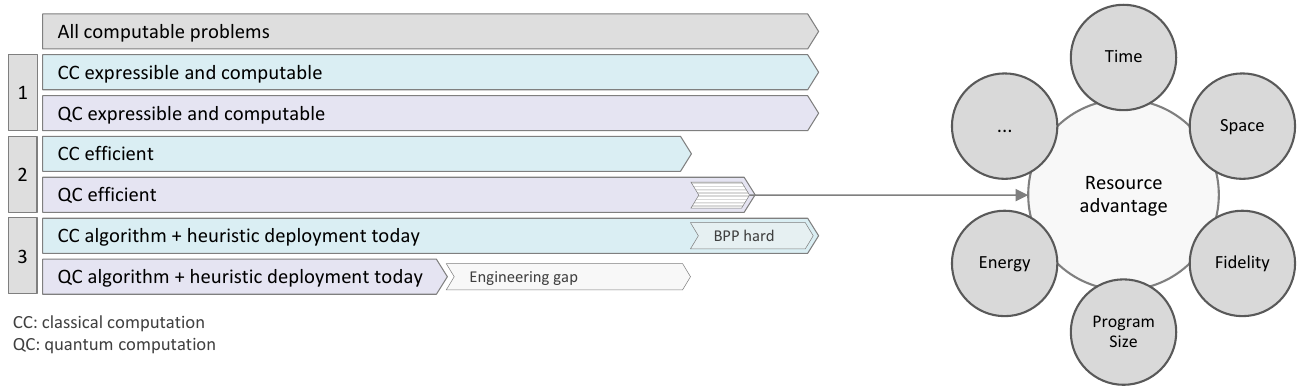}
    \caption{\centering Comparison of classical and quantum computation across (1) computability, (2) resource efficiency, and (3) pragmatic deployment. Quantum resource advantages are the differentiating factor in achieving provable quantum advantage, driving quantum accelerator research and development.}
    \label{fig:qadvantage}
\end{figure}

Computational resources are pivotal for quantifying the capabilities and efficiency of software and firmware in both classical and quantum paradigms. 
Quantum computation being in the same Turing degree of computability as classical computation, as shown in Figure~\ref{fig:qadvantage} (annotation 1); it is rather the frugality of required computational resources (annotation 2) that leads to an advantage in using a quantum accelerator in terms of efficiency.
This motivates a brief survey of computational resource metrics and their estimation techniques in this section.

Conventional computational resources are quantified in terms of time complexity, i.e., the measure and scaling of time needed for the algorithm; space complexity, i.e., the measure and scaling of data memory needed for the algorithm; and energy, i.e., the total heat incurred while running the computation. 
Two other valuable resources are approximation and description complexity. 
Approximation techniques in computation reduce time or space complexity by providing approximate solutions within a bounded error or infidelity. 
Description complexity is the measure of the amount of information contained in an object.
It can be expressed as the minimal length of a program that can perform the required computation, quantified by the upper semicomputable Kolmogorov complexity in a universal computing model such as a Turing machine. 

Time and space resources are typically expressed as asymptotic relations to the problem size.
Their (worst-case, average-case, and best-case) scaling complexity is usually quantified using Big-O, Big-$\Theta$, or Big-$\Omega(N)$ notations and expressed in units of cycles (or wall-clock time) for time and bits for memory. 
Additionally, these complexities depend on hardware specifics, as reflected in the algorithm's execution time and the additional temporary space required.
The energy cost can be measured in Joules (J) or kilowatt-hours (kWh), depending on the scale and context of the computation, using power-monitoring tools.
The description complexity $DC$ (also called Kolmogorov complexity, algorithmic complexity, or algorithmic entropy)~\cite{kolmogorov1965} of a string $s$ is formally defined as the length of the shortest program $p$, which, when fed into a prefix-free universal Turing machine $U$, computes the string $s$ as the output., i.e., $DC (s) = \min_p \{\|p\|: U(p) = s\}$.
Here, $\|p\|$ refers to the length of the program $p$ in bits. 
A key invariance property of $DC$ states that it is nearly independent of the choice of the universal automaton $U$, i.e., shifting from one language to another alters the complexity by only a constant that depends on the length of the language translation (e.g., a cross-compiler) and not the string itself. 
Although theoretically crucial, this complexity metric cannot be exactly computed due to limitations akin to the halting problem, but can be approximated using various techniques. 
Kolmogorov complexity, along with other closely related measures such as Martin Löf randomness, Chaitin's omega number~\cite{chaitin1966}, and Solomonoff's algorithmic probability~\cite{solomonoff_AP}, is studied in the field of algorithmic information theory~\cite{li2008introduction}.
Description complexity has also been defined in conjunction with other computational resources, such as time and approximation, to formulate compound resource-cost measures such as Levin complexity~\cite{levin1973universal}, Bennett's logical depth~\cite{Bennett1988-BENLDA-10}, and Schmidhuber's speed prior~\cite{speedprior-schmidhuber}.

\subsection{Quantum computational resources}
\label{qresources}

In the circuit/gate model of quantum computation, these metrics are mirrored by counterparts such as quantum circuit depth for time complexity and the number of qubits in the system for space complexity. 
These come under the purview of quantum complexity theory~\cite{aaronson2005complexity}.
Exemplary classes, like BQP (bounded-error quantum polynomial time), encapsulate the class of decision problems solvable in polynomial time scaling by a quantum Turing machine (QTM) (or a uniform family of polynomial-size quantum circuits~\cite{deutsch1989quantum}, using any universal gate set with efficiently computable transition amplitudes), with at most 1/3 probability of error.
BQP is synonymous with the class of feasible problems for quantum computers.

Pragmatically, the time cost can include additional subtleties, such as control parallelization enabled by the QPU, the time per gate operation, and auxiliary temporal overhead, such as error decoding or gate distillation.
Similarly, the space complexity needs to account for additional qubits for error correction or holes in Spin-qubit architectures for shuttling operation~\cite{paraskevopoulos2025arta}.
Moreover, space and time can be traded off using techniques like Bennett's pebbling game~\cite{bennet_tradeoff}.

\begin{figure}[htb]
    \centering
    \includegraphics[width=\textwidth,trim={1.5cm 0cm 0 1cm},clip]{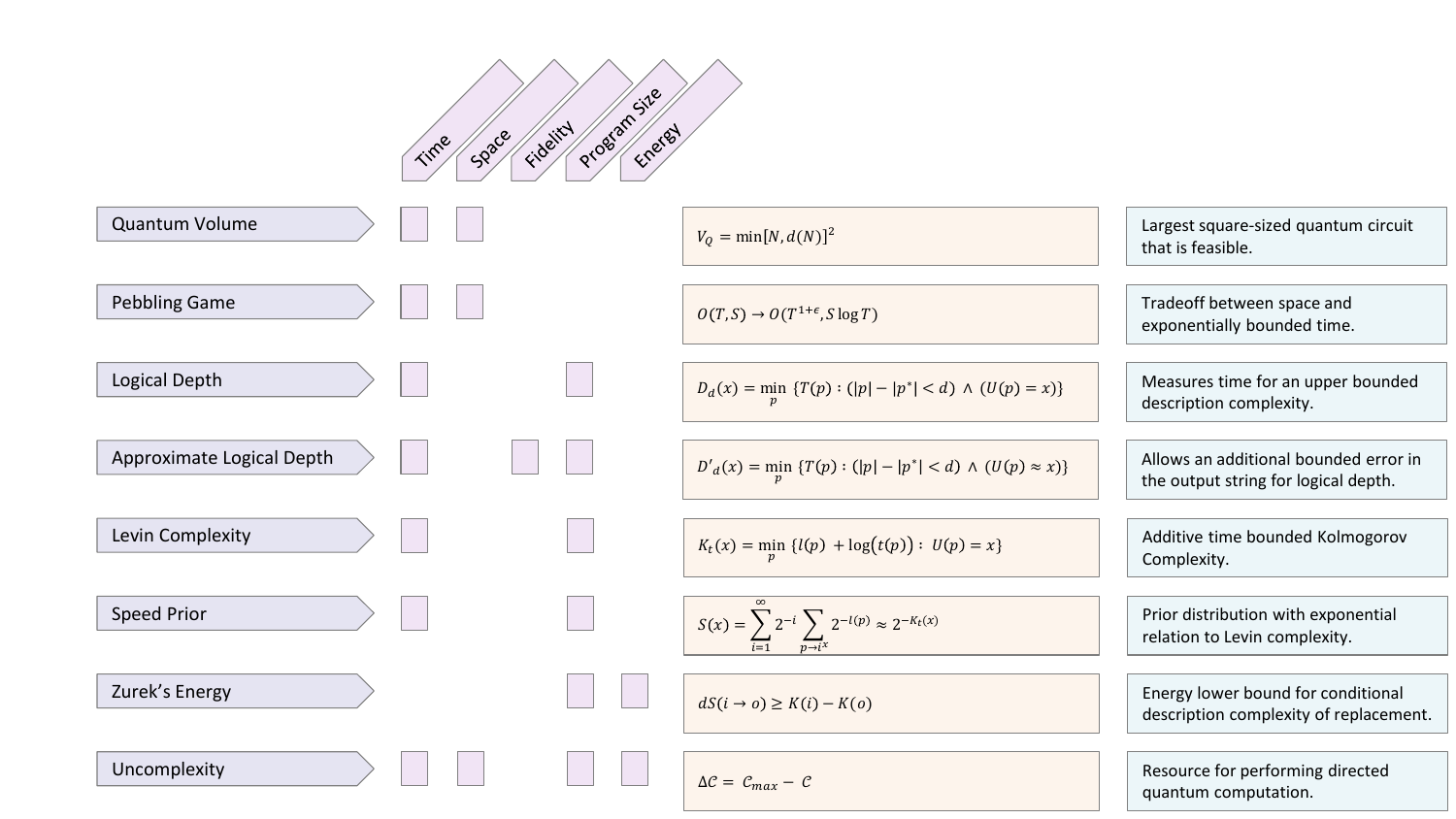}
    \caption{\centering Definitions of metrics that involve trade-offs between multiple resources - time, space, fidelity, program size, and energy.}
    \label{fig:resource_tradeoffs}
\end{figure}

Concepts of algorithmic information theory and description complexity can be extended to the scope of quantum computation via QTM \cite{deutsch_QTM}.
The notion of quantum Kolmogorov complexity for a quantum state refers to the minimum amount of either quantum~\cite{berthiaume2001quantum} or classical~\cite{vitanyi2001quantum} information needed to describe a program. 
Similar to its classical counterparts, it is only upper semi-computable and can thus only be approximated from above (i.e., a value higher than the actual complexity).
This program, when executed on a universal quantum computer, produces the desired quantum state with high accuracy.
The measures involving multiple resources, such as Levin complexity, logical depth, and speed prior, can similarly be extended to quantum computation, with applications in quantum intelligence agent models~\cite{sarkar2022qksa,sarkar2021estimating,sarkar2022applications}.
These metrics allow us to consider a holistic view of quantum computational resources from both theoretical and pragmatic perspectives.
The table in Figure~\ref{fig:resource_tradeoffs} summarizes the compound metrics.

Two metrics relate closely to the energy cost of quantum circuits.
Firstly, the thermodynamic cost of computation~\cite{zurek1989thermodynamic} proposes a correlation between the description complexity and energy costs in quantum computation. 
The author presents the mathematical foundation for establishing the minimal increase in entropy caused by a computational process that takes an input string $s$ to an output string $t$ as the conditional Kolmogorov complexity of the two strings.  
Secondly, the second law of quantum complexity~\cite{brown2018second} correlates the quantum complexity of a system with $K$ qubits and the evolution of the entropy of a classical system with $2^K$ degrees of freedom. 
The corresponding total entropy of a classical system is determined by two factors: the change in positional entropy, which parallels the evolution of quantum computational complexity, and the shift in kinetic entropy, which is analogous to the quantum Kolmogorov complexity. 
Thus, the total complexity $\mathcal{C}$ of a quantum program can be represented as a sum of the quantum circuit complexity (i.e., the quantum circuit depth multiplied by the number of qubits in the system) and the quantum description complexity. 
The circuit complexity of the quantum system for a physical system would be guided by the evolution of a Hamiltonian for the quantum system on a complexity geometry~\cite{dowling2006geometry}.
This circuit is then decomposed to simulate the system on a quantum computer, subject to connectivity and gate-set constraints.
The formulation in this article incorporates the approximation measure within the unitary decomposition to an available discrete universal gate set~\cite{sarkar2024yaqq}.

The estimation technique for the quantum Kolmogorov complexity is not explicitly addressed in \cite{brown2018second}.
Given the relationship between protocol complexity from \cite{kolchinsky2023generalized} and energy, the Kolmogorov complexity of the quantum program can be considered a measure of energy.
Hence, the total complexity forms a crucial multi-modal metric covering the resources of space, time, description complexity, and energy.
In this article, we provide an empirical measure of total complexity as the sum of the quantum volume and the compressed bit length of the EQISA representation.
In Section~\ref{application3_uncomplexity}, we use EQISA to estimate the total complexity of benchmark quantum algorithms to discern operational QC resources.
Fine-grained pulse-level energetic cost~\cite{fauquenot2025open} can also be incorporated and is left for future extension.

\subsection{Description of quantum computation}
\label{qdescription}

The information content of quantum programs can be expressed in different forms, each with a distinct quantum description. 
The quantum computing stack layers manifest in different representations at each level. 
Here, we examine different quantum descriptions and lay the foundation for choosing a unitary matrix as the quantum description.
\begin{enumerate}[nolistsep,noitemsep]
    \item Unitary matrix: Represents the transformation of a $n$-qubit quantum state in a closed quantum system with a $2^n \times 2^n $ matrix $U$ of complex entries satisfying $UU^\dagger = I$. 
    The unitary transformation can also be parameterized in terms of a Hamiltonian $\hat{H}$ of evolution for time $t$, as $U = e^{-i\hat{H}t}$. Open quantum systems are described by density matrices as states and Lindblad operators for the evolution.
    \item Quantum circuit: Provides a visual and operational representation consisting of logical units called quantum gates in a spatiotemporal circuit.
    \item Quantum assembly language (QASM): Textual representation \cite{openQASM, cQASM} that offers a human and machine-readable description of quantum circuits.
    \item Native-mapped QASM: Adaptation of the assembly-level quantum description according to the hardware-specific configurations, such as the native gate set and connectivity of qubits. 
    \item Quantum Instruction Set Architecture (QISA) microcode: Bit-level encoding of the native QASM  that is useful for designing the real-time control microarchitecture for interfacing with the arbitrary waveform generators \cite{eQASM}.
    \item Pulse control: Platform-specific signals, e.g., microwave or laser pulses, tuned for executing desired operations on target QPU platforms \cite{qiskit_pulse}.
\end{enumerate}

The unitary matrix for a quantum operation is significant as it captures the complete information of the evolution of the quantum system in a succinct manner. 
The decomposition of unitary matrices into sequences of quantum gates is essential for translating abstract quantum algorithms into executable instructions for quantum hardware. 
This decomposition bridges the continuous-discrete divide inherent in quantum computations, moving from the continuous evolution described by unitary operators in $SU(n)$ to discrete gate sequences suitable for practical implementation.

\subsection{Estimation of description complexity}
\label{approxdescription}

In the preceding section, we discussed description complexity as a pivotal measure in quantifying the intrinsic information content of a data object. 
Despite its theoretical importance as a resource measure in computation, a fundamental obstacle to Kolmogorov complexity is its non-computability: no algorithm exists that, given a string, produces its exact Kolmogorov complexity as output. 
This limitation arises because calculating the true Kolmogorov complexity entails solving the halting problem, which is undecidable \cite{Turing_1937}.
Nonetheless, the inability to compute Kolmogorov complexity precisely does not diminish its utility in theoretical and applied informatics. 
Instead, it necessitates exploring viable methods for estimating this complexity via its upper semi-computability. 
For example, the block decomposition method (BDM) \cite{bdm_zenil} estimates the algorithmic complexity of smaller blocks of strings (with Levin's coding theorem method (CTM) \cite{levin1974laws}), which are then summed up to reconstruct the complexity of the original data. 
BDM offers a hybrid approach to evaluating complexity by combining Shannon entropy at the long-range scale with local predictions of algorithmic complexity.
Although BDM provides a sharper estimate of description complexity, it is not suitable for our purposes as it does not produce an explicit description or encoding that realizes the estimated complexity.

Another effective approximation technique for discerning Kolmogorov complexity is lossless data compression.
Compression serves as a practical proxy~\cite{zenil2021compression} by leveraging the fact that the size of the compressed data reflects the underlying patterns and redundancies, providing a tangible measure of the data's complexity.
Lossless compression exploits statistical regularities while incurring no loss of information. 
Standard lossless compression methods include bzip2~\cite{seward1996bzip2}, LZ~\cite{LZ77}, and minimal description length (MDL)~\cite{rissanen1978modeling}.
bzip2 uses the Huffman prefix coding technique, which is part of this research.
The bzip2 pipeline consists of 5 steps: (i) run length encoding 1, (ii) Burrows-Wheeler transform (BWT) \cite{burrows1994block}, (iii) move-to-front transform (MTF), (iv) run length encoding 2, and (v) prefix coding.
bzip2 is very efficient at compressing text with repeated patterns and is asymmetric; decompression is faster than compression. 
As explored further in Section~\ref{bzip}, these qualities make it a suitable choice for the compression of QASM instructions of quantum circuits, which contain repeated instances of gate operations and qubit IDs. 


\subsection{Coding via sparse dictionary learning}
\label{coding}

Besides compression algorithms, efficient instruction coding can aid in compressing the information.
More generally, coding changes the form of information to make it more convenient for performing specific operations, such as storage, communication, error correction, or encryption. 
Variable-length codes, e.g., Morse codes, are widely used in lossless compression techniques and are essential to entropy encoding techniques. 
Entropy encoding methods prioritize frequently used symbols or characters with shorter codes, while assigning longer codes to less frequently used ones. 
This attribute is significant for data compression. 
Encoding the stream of instructions with variable-length encoding is reasonably straightforward, but the process becomes more involved when decoding~\cite {david_salomon}. 
The choice of encoding must be made with special consideration to ensure unique decoding, which is a prerequisite for lossless compression. 
This special consideration is the prefix property.
Prefix-free or simply prefix refers to the property of a set of symbols or strings such that no element of the set is a prefix of another member of the set.
Similarly, a prefix code is a coding scheme in which no code word is a prefix of another code word in the set. 

Huffman codes~\cite{huffman} extend this concept by constructing the most efficient compression for a given set of frequencies.
Given a source stream of symbols, it tabulates prefix codes of variable length for each constituent symbol based on the frequency of occurrence of the symbol. 
It works by creating a binary tree of nodes, called the Huffman tree, where each symbol gets a unique binary code, with more frequently occurring symbols placed closer to the root and assigned shorter codes based on the traversal from the root.
Huffman coding serves as the basis for the proposed EQISA in this article.

A dictionary coder is an implementation of a data compression algorithm that operates by searching for matches between the text to be compressed and a set of strings contained in a dictionary data structure maintained by the encoder. 
When the encoder finds such a match, it substitutes a reference to the string's position in the data structure.
Simple dictionary coders use a static dictionary whose full set of strings is determined before coding begins and remains unchanged throughout the coding process. 
More common are methods in which the dictionary starts in a predetermined state but its contents change during the encoding process based on previously encoded data, e.g., in the LZ77~\cite{ziv2003universal} and LZ78~\cite{ziv2003compression} algorithms.

The flexibility and information efficiency of dictionary coding can be further enhanced by sparse dictionary learning. 
Sparse dictionary learning~\cite{tillmann2014computational} infers a sparse representation of the input data in the form of a linear combination of basic elements as well as those basic elements themselves. 
These elements, called atoms, compose the dictionary and, often, an overcomplete spanning set.
A dictionary trained to fit the input data can significantly improve sparsity.
This has applications in data decomposition, compression, and analysis, and has been used in image denoising and classification, compressed sensing, signal recovery, video and audio processing, and machine learning~\cite{agrawal2026single}. 
The novelty of the proposed EQISA in this article rests on inferring a sparse dictionary empirically that generalizes across arbitrary quantum unitaries and benchmark algorithms.


\section{Instruction set architecture}
\label{isa}

\begin{figure}[htb]
    \centering 
    \includegraphics[clip, trim=6.1cm 1.0cm 1.0cm 36.2cm, width = 1\linewidth]{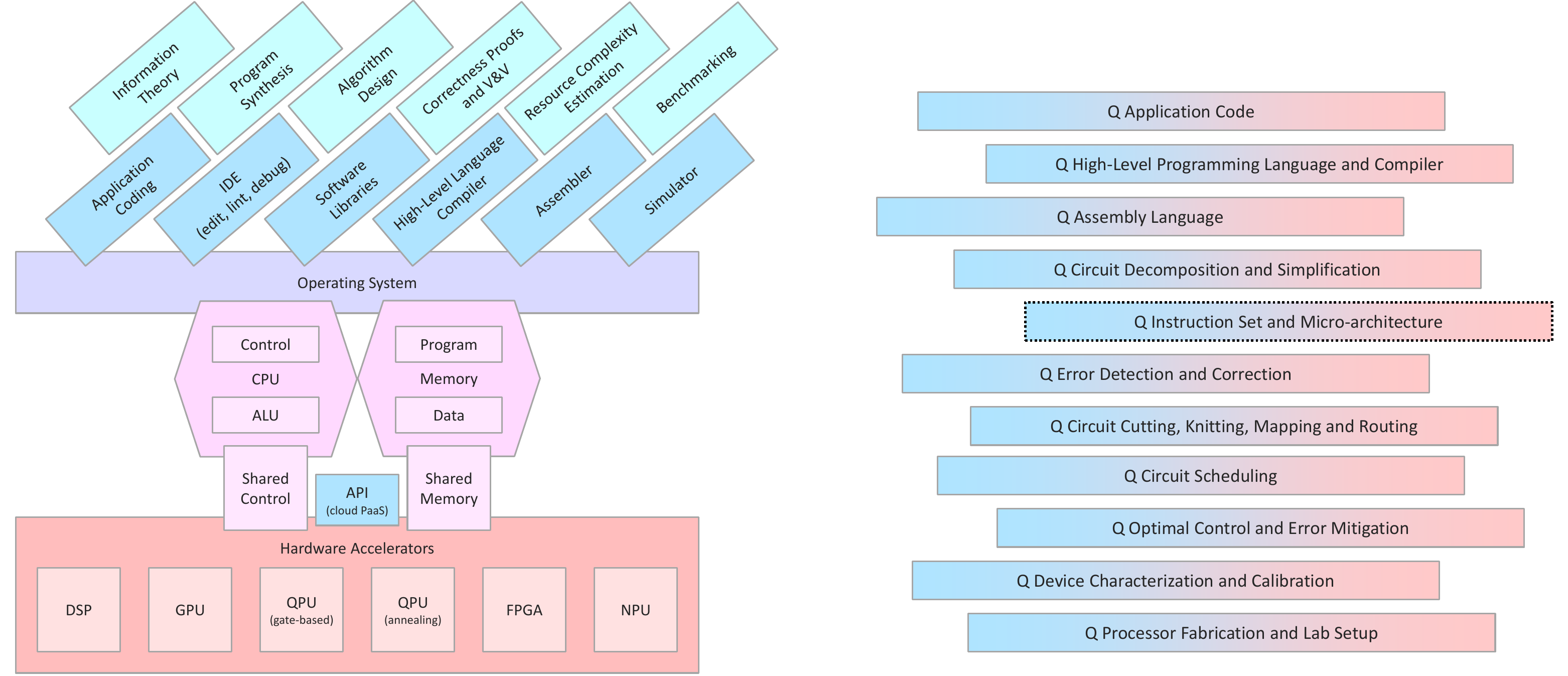}
    \caption{Overview of the system design of a quantum accelerator with classical control and various software modules required for research and development is shown on the left. The different abstraction layers for full-stack quantum computing are shown on the right. This research pertains to the quantum instruction set (indicated in a dotted outline).}
    \label{fig:qcstack}
\end{figure}

The interface between the QC applications and the QPU is organized into abstraction layers, called the quantum computation stack~\cite{bertels2020quantum,bertels2021quantum}, as shown in figure~\ref{fig:qcstack}.
The system design of a quantum accelerator with classical control and various auxiliary software modules is shown on the left, while the different abstraction layers for full-stack quantum computing are shown on the right.
In the QC stack, from top to bottom, first, the application, formulated as a quantum algorithm, is expressed in a quantum programming language. 
A quantum compiler thereafter decomposes and optimizes the high-level code into native operations supported by the target QPU. 
Then the quantum microarchitecture schedules and issues low-level instructions in real time. 
These instructions (such as initialization, unitary gates, and measurements) also need to be translated into corresponding analog pulses that optimally control the accessible degrees of freedom of the quantum system. These electromagnetic analog signals perform the necessary transformation for synthesizing quantum unitary gates on specific addressable qubits while mitigating the undesirable effects of noise. Eventually, these hardware-aware signals implement the desired logical operation dictated by the hardware-agnostic quantum algorithm on the target QPU.
This research pertains to the quantum instruction set outlined in dotted lines.

An instruction set architecture (ISA) is an abstract model that specifies the programmable interface of a computer and how software interacts with hardware.
In general, an ISA includes the instructions, data types, registers, addressing modes, virtual memory, etc, and how each is specified, for example, as opcodes.
ISA differs from a microarchitecture in that the latter concerns itself with processor design and the control flow for interpreting the ISA. 

\subsection*{Compressed instruction set architecture}

Code compression has been applied to low-power embedded systems. 
Low-power embedded systems are computing systems designed for specific functions with a primary focus on minimizing energy consumption. 
These systems are integral to a wide range of applications, from consumer electronics to industrial control systems, where power efficiency is critical. 
Code-compression techniques are advantageous for overcoming the energy, size, and memory constraints that limit such systems.

The study of the design space of code compression~\cite{MENON2014179} is an active field of research that involves choosing between statistical and dictionary-based encoding techniques and minimizing decompression overhead, hardware cost, and power usage. 
The RISC-V ISA~\cite{risc-manual} is an open-source ISA for standard and special-purpose utility and operates with fixed-length 32-bit instructions. 
It has many extensions for applications in embedded systems, personal computers, supercomputers, and quantum computation. 
Approaches like \cite{RISC-V} compare the performance of some of these extensions and explore the design space of compressed ISA by employing arithmetic encoding to assign shorter-length opcodes to more commonly used instructions and increase the code density.
Specifically, the RISC extension RVC~\cite{waterman2011} employs variable-length opcodes to reduce instruction bandwidth.
We apply a similar technique as RVC for quantum instructions.

Extending the principles of code compression and efficient encoding to the domain of quantum computation is meaningful for multiple reasons. 
Investigation of compression in quantum computation provides a strategy for estimating the description complexity of quantum circuits. 
Secondly, attempts to move quantum control architectures closer to the operating temperatures of quantum processors impose severe constraints on the energy budget. 
Cryogenic control of qubits offers better integration with the processor and shorter process times, eventually qualifying for a promising scalable alternative for quantum control \cite{cryo_fpga}. 
Increasing code density can reduce the energy cost of information transfer from room temperature to cryogenic temperatures, thereby lowering operating power and improving noise performance.
This application is further demonstrated via our proposed EQISA in Section~\ref{application1_energy_gains}.

\section{Unitary decomposition}
\label{ud}

A set of quantum logic gates is considered computationally universal if any quantum computation can be efficiently expressed up to error-bounds using those gates.
The unitary matrix, $U$, can be decomposed into $k$-local gates, i.e., using gates that act on at most $k$ qubits.
Note that this is independent of the physical locality of the QPU's qubits.
The routing process ensures the physical locality of the $k$ qubits involved in these $k$-local gates and introduces a worst-case constant-factor overhead~\cite{steinberg2024lightcone} to the runtime, which depends on the QPU size. 
For example, the Deutsch gate $D(\theta)$ with $k=3$ is a single-parameter gate that is universal for QC.
Universal QC can be achieved~\cite{lloyd1995almost} with $k=2$, such as the $2$-qubit CX gate and single-qubit arbitrary angle rotation gate along any $2$ mutually orthogonal axes.
It has been proven that an exact decomposition of an arbitrary $n$-qubit gate requires at least $\frac{1}{4}(4n-3n-1)$ CX gates~\cite{krol2024highly,shende2004minimal}.
This worst-case exponential cost for $k$-local universal QC implies that only a small subset of unitaries will be practical for expressing and executing quantum algorithms.
The corresponding algorithm to decompose $U$ is the quantum logic counterpart~\cite{shende2005synthesis} of Shannon decomposition of a Boolean function~\cite{shannon1949synthesis}.
Quantum Shannon decomposition (QSD) is an important first-order decomposition that can easily be recursively generalized to $n$-qubit.
The algorithm is explained further in Appendix~\ref{appendix:qsd}.

The single-qubit arbitrary rotations need to be further decomposed into a discrete gate set, typically Hadamard and T gates, which is universal for 1-qubit computation.
Though current QPU models support arbitrary rotations along X, Y, and Z axes for physical qubits, in this work, we consider a finite set of discrete gates for the decomposition.
We justify this choice threefold. 

Firstly, quantum characterization is exponentially resource-intensive.
Full characterization, for example, via gate-set tomography~\cite{yu2024transformer}, is typically performed for a small set of rotation angles (e.g., $90^\circ$ and $45^\circ$ for each of the $3$ axes).
The characterization is, in turn, used to tune the control electronics and the compilation process; thus, only the subset of gates characterized can be reliably used in the QPU.

Secondly, large-depth fault-tolerant quantum computation (FTQC) necessitates employing QEC, which encodes the quantum information of each logical qubit using a set of physical qubits.
This encoding then performs a universal set of operations at the logical level by local operations on the physical level.
Some operations that can be easily translated to this local form while maintaining the fault tolerance of the QEC code are termed transversal gates.
Transversality is typically proven for specific discrete gates using a specific QEC code rather than a family of parametric gates.
Moreover, due to the Eastin-Knill theorem~\cite{eastin2009restrictions}, it is known that transversal gate sets cannot be universal and require additional resources, such as magic-state distillation.
Thus, FTQC will be composed of a small set of transversal gates for the chosen QEC code and additional resource states.
Therefore, every unitary matrix $U$ must be decomposed into this set of discrete operations with maximum fidelity before introducing additional resource states.

Thirdly, and most relevant to this work, in a quantum control setting, we assume that the QC is operated for arbitrary parametric gates that require a continuous parametric space~\cite{sayginel2023fault}.
However, the precision of these parameters is typically discretized by the quantum programming language's datatype encoding the angle, the microarchitecture's quantum instruction bandwidth, and the digital-to-analog converter's resolution for microwave pulse control.
While the set of discrete controllable gates is considerably large in this setting, as the size of the Hilbert space grows with larger quantum systems, the reachable volume of the Hilbert space within bounded errors will still exponentially reduce, considering the $k$-locality of these gates from QSD.

The Solovay-Kitaev decomposition~(SKD)\cite{dawson_nielsen}  allows one to decompose an arbitrary $1$-qubit $U$ using the discrete gate set that forms a dense subgroup of $SU(2)$ and is closed under inversion, e.g., [$H$, $T$, $T^\dagger$].
In its general form, the corresponding Solovay-Kitaev theorem (SKT) states that for a finite set of elements $\mathcal{G}$ (drawn from a special unitary group $SU(2)$ containing its inverses that generates a dense group) and precision error $\epsilon > 0$; for any $U \in SU(2)$ there is a sequence $S$ of gates in $\mathcal{G}$ of length $O(\log^{\log(5)/\log(3/2)}(1/\epsilon))$ (Equation~8 in \cite{dawson2005solovay}) such that the operator norm error is bounded $||S-U|| \le \epsilon$.
Moreover, the decomposition can be computed in $O(\log^{\log(3)/\log(3/2)}(1/\epsilon))$ time (Equation~9 in \cite{dawson2005solovay}).
$\mathcal{G}$ denotes the group generators (a finite subset of $SU(d)$ for $d$-dimensional qudit)~\cite{ozols2009solovay}.
The corresponding generators of the Lie algebra, $\mathfrak{su}(d)$, are the Hermitian matrices denoting the Hamiltonians for these generators (e.g., the Pauli matrices for $\mathfrak{su}(2)$ and the Gell-Mann matrices for $\mathfrak{su}(3)$). 
The implementation of SKD includes a preprocessing step that generates a search space of composite sequences up to length $l_0$ composed of gates from a finite discrete basis. 
This search space is referred to as the Solovay-Kitaev (SK) basis in the remainder of the article. 
The maximum length of sequences $l_0$ is referred to as the depth $d$ of the SK basis. 
The chosen set of fundamental gates and the group generated by the set, that is, the SK basis, must fulfill the following conditions generalized for an $m$-qubit system:

\begin{enumerate}[nolistsep,noitemsep]
    \item All the gates in the set belong to the group of special unitary matrices $SU(m)$ and have a determinant $1$.
    \item The set of gates is closed under inversion, implying that for every gate in the set, its hermitian conjugate must also belong to the set.
    \item The group that is generated by the set must densely span the space $SU(m)$. This means that, for every arbitrary unitary operation $U$, there must exist a product sequence of gates from the set that can approximate $U$ with a bounded error $\epsilon$.
\end{enumerate}

\begin{figure}[htb]
    \centering
\pgfdeclarelayer{background}
\pgfsetlayers{background,main}
\tikzstyle{startstop} = [circle, minimum width=0.5cm, minimum height=0.5cm,text centered, draw=teal, fill=green!10]
\tikzstyle{io} = [trapezium, rounded corners, trapezium left angle=70, trapezium right angle=70, minimum width=1cm, minimum height=0.5cm, text centered, draw=violet, fill=violet!10]
\tikzstyle{process} = [rectangle, rounded corners, minimum width=1cm, minimum height=0.5cm, text centered, draw=blue, fill=blue!10]
\tikzstyle{decision} = [diamond, minimum width=1cm, minimum height=0.5cm, text centered, draw=orange, fill=yellow!30]
\tikzstyle{arrow} = [thick,->,>=stealth]

\begin{tikzpicture}[node distance=1.8cm]
    \small
    \node (start) [startstop] {\texttt{Start}};
    \node (initiate) [process, below of=start, yshift=+0.2cm, text width=4cm] {\texttt{initiate Solovay\_Kitaev (U,n)}};
    \node (check) [decision, below of=initiate, yshift=-0.5cm, text width=1.5cm] {\texttt{check if n == 0}};
    \node (skcall) [process, below of=check, yshift=-0.6cm] {$U_{n-1}$ = \texttt{Solovay\_Kitaev(U, n-1)}};
    \node (decompose) [process, below of=skcall, yshift=+0.7cm] {$V_n$, $W_n$ = \texttt{balanced\_commutator\_decompose(}$UU^{\dagger}_{n-1}$\texttt{)}};
    \node (skv) [process, below left of=decompose, xshift=-2.5cm] {$V_{n-1}$\texttt{ = Solovay\_Kitaev(}$V_n$\texttt{, n-1)}};
    \node (skw) [process, below right of=decompose, xshift=2.5cm] {$W_{n-1}$\texttt{ = Solovay\_Kitaev(}$W_n$\texttt{, n-1)}};
    \node (return) [io, below of=decompose, yshift=-0.6cm] {\texttt{return} $U_{n} = V_{n-1}W_{n-1}V_{n-1}^{\dagger}W^{\dagger}_{n-1}U_{n-1}$};
    \node (bestapprox) [io, left of=check, xshift=-2.5cm, text width=3.2cm] {\texttt{return best\_approximation of U}};
    
    \draw [arrow] (start) -- (initiate);
    \draw [arrow] (initiate) -- (check);
    \draw [arrow] (check) -- node[anchor=east] {False} (skcall);
    \draw [arrow] (check) -- node[anchor=south] {True} (bestapprox);
    \draw [arrow] (skcall) -- (decompose);
    \draw [arrow] (decompose) -- (skv);
    \draw [arrow] (decompose) -- (skw);
    \draw [arrow] (skw.east) -- ++(0.5,0) |- (initiate.east);
    \begin{pgfonlayer}{background}
    \draw [arrow] (skv.west) -- ++(-0.5,0) |- (initiate.west);
    \end{pgfonlayer}
    \draw [arrow] (skv) -- (return);
    \draw [arrow] (skw) -- (return);
    
    \end{tikzpicture}
    \caption{Flowchart of the Solovay-Kitaev decomposition algorithm for 1-qubit unitary quantum operator $U$ and recursion depth $n$. It returns the $\epsilon_n$ approximation to the target unitary $U$ computed through the function call at the $n-1$ degree of recursion, and returns the $\epsilon_0$ approximation in the base case.}
    \label{tikz: flowchart} 
\end{figure}
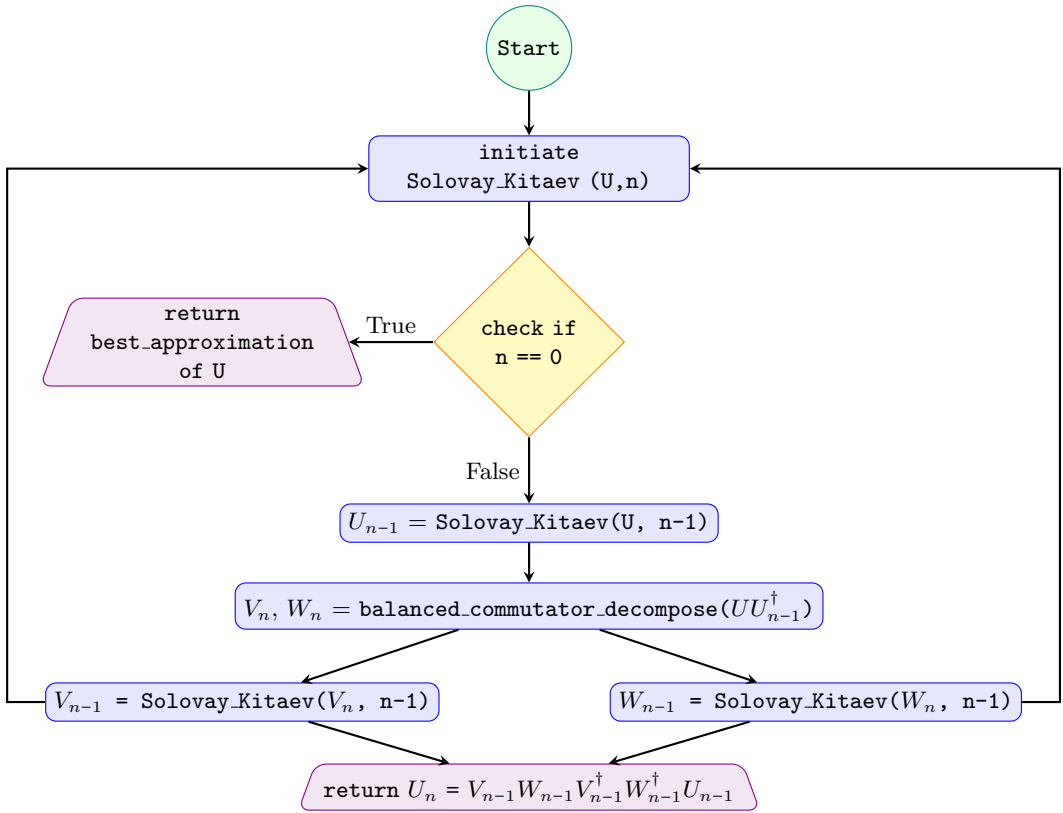


We examined the implementation of SKD in Qiskit to reconstruct the sequence rearrangement at each level of progression from the base case.
As shown in the Figure~\ref{tikz: flowchart}, the algorithm functions in a recursive fashion, and the degree of recursion is denoted by $n$ in the flowchart. 
Throughout the process, the algorithm returns a sequence that approximates the unitary operator $U$ to within an error of $\epsilon_n$. The approximation error at each recursion level $r$ is related to that at the level $(r-1)$. This goes on until the base case that returns the \texttt{best\_approximation} to matrix $U$, which is bounded by $\epsilon_0$. 

The algorithm is designed to obtain an improved approximation accuracy $\epsilon_r < \epsilon_{r-1}$. 
The approximation accuracy tends to $0$ as the recursion depth $n$ increases. 
The \texttt{balanced\_commutator\_decompose} method performs a balanced group commutator decomposition of the accuracy at level $r$ defined as $\Delta = UU_{r-1}^\dagger = VWV^\dagger W^\dagger$ for matrices $V$ and $W$. $(r-1)$ level approximation accuracies are computed by the call of the function again for matrices $V$ and $W$ and the $r^\text{th}$ level approximate sequence $U_r$ = $V_{r-1}W_{r-1}V_{r-1}^\dagger W_{r-1}^\dagger U_{r-1}$, consisting of all $5$ terms computed from the $(r-1)^\text{th}$ level is returned.

\section{Method and benchmarking}
\label{method}


In this section, we will present the proposed encoding methods used within EQISA.
Thereafter, the methods are evaluated against random and algorithmic datasets.
Finally, we augment the proposed sparse dictionary learning variant with the bzip2 lossless compression as the proposed pipeline.

\subsection{QISA encoding techniques}
\label{sec_4.2}

Just as lexical constructs called words are sequentially composed of individual units called characters, so too are quantum circuits systematically constructed through the sequential application of quantum operations. 
These quantum logic gate operations serve as fundamental units for building the quantum circuit as sequences of gates (as represented in Figure~\ref{fig:skd_circ}). 
The subsequent analysis presents different encoding versions that increasingly examine and harness structure within the fundamental components of quantum circuits. 
The gate set [$H$, $T$, $T^\dagger$] has been consistently selected for all analyses in the work.

\subsubsection{Binary encoding (v0)}

To establish a baseline for encoding efficiency, we implement a uniform binary encoding scheme for quantum circuits. 
Each gate within the circuit is represented by a fixed-length binary code. 
The length of each code, denoted as $b$, is calculated based on the total number of distinct gates in the dictionary as $b = \lceil \log_2(N) \rceil$, where $N$ is the number of distinct gates in the gate-set. 
This approach ensures that each gate is uniquely representable and uses the minimum number of bits required for that representation. 
For the gate set [$H$, $T$, $T^\dagger$], we need $2$ bits per gate. The total information content of the circuit, measured in bits, is the product of $b$ and the total number of gates in the circuit. 

As an example, consider a Harr-random matrix denoted as $U$ prior to undergoing decomposition in terms of the gate set [$H$, $T$, $T^\dagger$]:
\begin{equation}
    U = 
    \begin{bmatrix}
    0.50359966+0.62609046j & -0.07233711+0.59090224j \\
    0.31138773+0.50738132j &  0.19782201-0.77876077j
    \end{bmatrix}
    \label{unitary matrix}
\end{equation}

\begin{figure}[htb]
    \centering
    \includegraphics[width=0.8\textwidth]{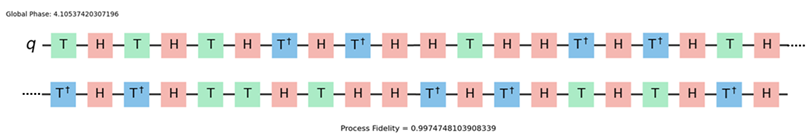}
    \caption{\centering Exemplary quantum circuit for a Haar-random sampled unitary (Equation~\ref{unitary matrix}) decomposed via Solovay-Kitaev decomposition with recursion degree $n=2$ and depth $d=3$.}
    \label{fig:skd_circ}
\end{figure}

The resulting decomposed quantum circuit, achieved via Solovay-Kitaev decomposition with a recursion degree $n=2$ and depth $d=3$, is depicted in Figure~\ref{fig:skd_circ}.

This circuit has a depth of $40$, which implies that it needs $40 \times 2 = 80$ bits of information using the v0 encoding. 

\subsubsection{Huffman encoded gate set (v1)}
\label{huff_v1}

The first version of Huffman coding for decomposed quantum circuits uses the gates in the chosen gate set as the building blocks of the quantum circuit. 
The gates are assigned codes based on how many times they appear in a circuit.  

In the example of the unitary matrix of Equation~\ref{unitary matrix} and its decomposed quantum circuit in Figure~\ref{fig:skd_circ}, we count the gate frequencies.
As shown in Figure~\ref{fig:tree huff v1}, the frequency distribution is used to construct the Huffman tree, and thereby the Huffman code.
The number of bits needed for describing the quantum circuit using this version of Huffman coding is, therefore, $21\times1 + 10\times2 + 9\times 2 = 59$. 
This is a notable improvement over binary encoding, which needs $80$ bits. 
The compression factor for this particular decomposition is $59/80 = 0.7375$.

\begin{figure}[htb]
    \centering
    \begin{minipage}[t]{.30\textwidth}
    \vspace{0pt}
    \centering
        \begin{tabular}{c|c}
            Gate & Frequency \\
            \hline
            $H$ & 21 \\
            $T$ & 10 \\
            $T^\dagger$ & 9 
        \end{tabular}
    \end{minipage}
    \hfill
    \begin{minipage}[t]{.30\textwidth}
    \vspace{0pt}
    \centering
        \begin{forest}
        for tree={
            draw,
            minimum size=2.5em,
            s sep=8mm
          },
          inner/.style={circle},  
          leaf/.style={rectangle, minimum width=2em, fill=gray!30},  
          EL/.style={edge label={node[midway,left,font=\scriptsize]{0}}},
          ER/.style={edge label={node[midway,right,font=\scriptsize]{1}}} 
          [40, inner
            [19, inner, EL
                [$T^\dagger$, leaf, EL]
                [$T$, leaf, ER]
                ]
            [$H$, leaf, ER]
            ]
        \end{forest}
    \end{minipage}
    \hfill
    \begin{minipage}[t]{.30\textwidth}
    \vspace{0pt}
    \centering
        \begin{tabular}{c|c}
            Gate & Huffman Code \\
            \hline
            $H$ & 1 \\
            $T$ & 01 \\
            $T^\dagger$ & 00 
        \end{tabular}   
    \end{minipage}

    \begin{subfigure}[t]{.30\textwidth}
    \caption{Frequency distribution of gates in the decomposed circuit.}
    \end{subfigure}
    \hfill
    \begin{subfigure}[t]{.30\textwidth}
    \caption{Huffman tree representation for encoding quantum gates.}
    \end{subfigure}
    \hfill
    \begin{subfigure}[t]{.30\textwidth}
    \caption{Huffman code table corresponding to adjoining tree.}
    \end{subfigure}
    \caption{\centering Huffman v1 encoded gate set for the example circuit.}
    \label{fig:tree huff v1}
\end{figure}


\begin{figure}[hbt]
    \centering
    \includegraphics[width=\textwidth, clip, trim={1.2cm 0cm 0.8cm 7.5cm}]{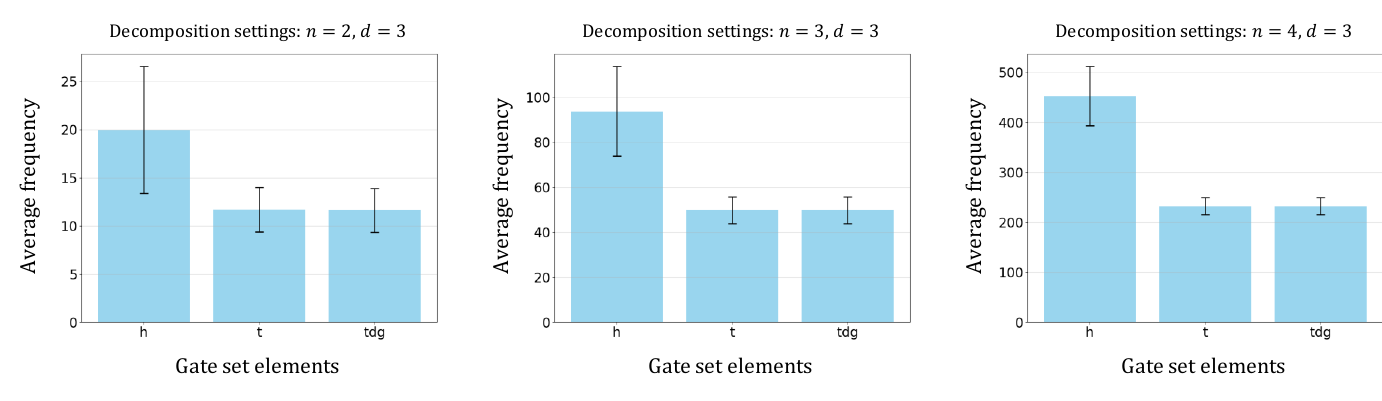}
    \caption{Bar charts showing the average distribution of gate frequencies [$H$, $T$, $T^\dagger$] in decomposed circuits obtained from Solovay-Kitaev decomposition of the data set of unitaries with $d=3$ and $n=2, 3, 4$.}
    \label{fig:avg-gateset-lists}
\end{figure}

To assign a general code to the gates, a dataset of $200$ Haar-random unitaries was selected. 
Following the Solovay-Kitaev decomposition with a depth $d=3$ and varying the degree of recursion $n = 2, 3, 4$, the gate frequencies in the decomposed circuits were stored.
This distribution of frequencies is observed consistently across varying choices of $n$ as shown in Figure~\ref{fig:avg-gateset-lists}. 
This allows the adoption of a general Huffman code to encode the decomposed quantum circuit for any random unitary matrix.

\subsubsection{Huffman encoded SK basis (v2)}
\label{huff_v2}

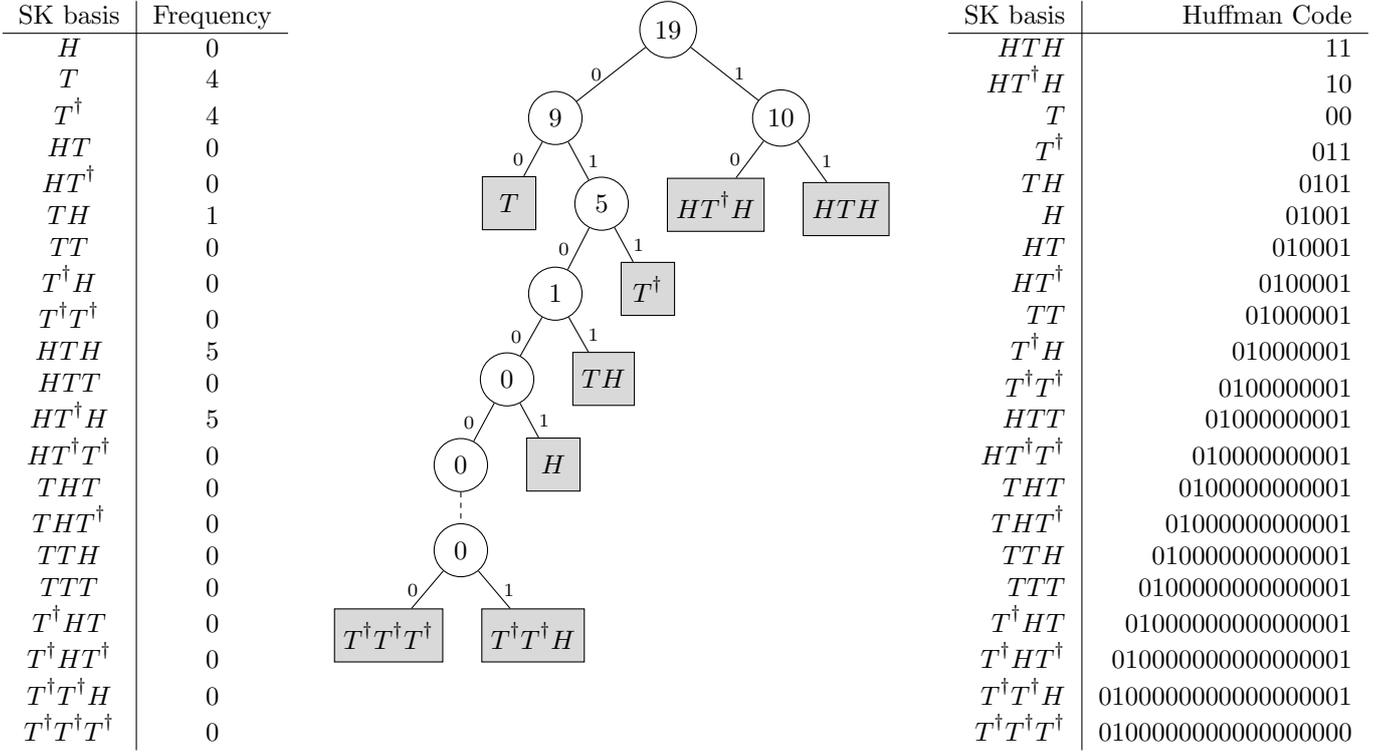
\begin{figure}[!bht]
    \centering
    \begin{minipage}[t]{.2\textwidth}
    \vspace{0pt}
    \centering
        \renewcommand{\arraystretch}{1.0}
        \begin{tabular}{c|c}
            SK basis & Frequency \\
            \hline
            $H$ & 0 \\
            $T$ & 4 \\
            $T^\dagger$ & 4 \\
            $HT$ & 0 \\
            $HT^\dagger$ & 0 \\
            $TH$ & 1 \\
            $TT$ & 0 \\
            $T^\dagger H$ & 0 \\
            $T^\dagger T^\dagger$ & 0 \\
            $HTH$ & 5 \\
            $HTT$ & 0 \\
            $HT^\dagger H$ & 5 \\
            $HT^\dagger T^\dagger$ & 0 \\
            $THT$ & 0 \\
            $THT^\dagger$ & 0 \\
            $TTH$ & 0 \\
            $TTT$ & 0 \\
            $T^\dagger HT$ & 0 \\
            $T^\dagger HT^\dagger$ & 0 \\
            $T^\dagger T^\dagger H$ & 0 \\
            $T^\dagger T^\dagger T^\dagger$ & 0
        \end{tabular}
    \end{minipage}
    \begin{minipage}[t]{.48\textwidth}
    \vspace{0pt}
    \centering
        \begin{forest}
        for tree={
            draw,
            minimum size=2em,
            s sep=5mm
          },
          inner/.style={circle},  
          leaf/.style={rectangle, minimum width=2em, fill=gray!30},  
          EL/.style={edge label={node[midway,left,font=\scriptsize]{0}}},
          ER/.style={edge label={node[midway,right,font=\scriptsize]{1}}}, 
          dotted/.style={edge={dotted, dash pattern=on 2pt off 2pt}}
          [19, inner
            [9, inner, EL
                [$T$, leaf, EL]
                [5, inner, ER
                    [1, inner, EL
                        [0, inner, EL
                            [0, inner, EL
                                [0, inner, dotted
                                    [$T^\dagger T^\dagger T^\dagger$, leaf, EL]
                                    [$T^\dagger T^\dagger H$, leaf, ER]
                                ]
                            ]
                            [$H$, leaf, ER]
                        ]
                        [$TH$, leaf, ER]
                    ]
                    [$T^\dagger$, leaf, ER]
                ]
            ]
            [10, inner, ER
                [$HT^\dagger H$, leaf, EL]
                [$HTH$, leaf, ER]
            ]
          ]     
        \end{forest}
    \end{minipage}
    \begin{minipage}[t]{.30\textwidth}
    \vspace{0pt}
    \centering
        \renewcommand{\arraystretch}{1.0}
        \begin{tabular}{r|r}
            SK basis & Huffman Code \\
            \hline
             $HTH$         &                    11 \\
             $HT^\dagger H$       &                    10 \\
             $T$                   &                    00 \\
             $T^\dagger$                 &                   011 \\
             $TH$              &                  0101 \\
             $H$                   &                 01001 \\
             $HT$              &                010001 \\
             $HT^\dagger$            &               0100001 \\
             $TT$              &              01000001 \\
             $T^\dagger H$            &             010000001 \\
             $T^\dagger T^\dagger$          &            0100000001 \\
             $HTT$         &           01000000001 \\
             $HT^\dagger T^\dagger$     &          010000000001 \\
             $THT$         &         0100000000001 \\
             $THT^\dagger$       &        01000000000001 \\
             $TTH$         &       010000000000001 \\
             $TTT$         &      0100000000000001 \\
             $T^\dagger HT$       &     01000000000000001 \\
             $T^\dagger HT^\dagger$     &    010000000000000001 \\
             $T^\dagger T^\dagger H$     &   0100000000000000001 \\
             $T^\dagger T^\dagger T^\dagger$   &   0100000000000000000 
        \end{tabular}   
    \end{minipage}

    \begin{subfigure}[t]{.27\textwidth}
    \caption{Frequency distribution of SK basis in the decomposed circuit.}
    \end{subfigure}
    \hfill
    \begin{subfigure}[t]{.37\textwidth}
    \caption{Huffman tree representation for encoding the Solovay-Kitaev basis.}
    \end{subfigure}
    \hfill
    \begin{subfigure}[t]{.30\textwidth}
    \caption{Huffman code table corresponding to adjoining tree.}
    \end{subfigure}
    \caption{\centering Huffman v2 encoded SK basis for the example circuit.}
    \label{fig:tree huff v2}
\end{figure}

The second version of Huffman coding for the decomposed quantum circuits leverages gate sequences derived from the Solovay-Kitaev (SK) basis as the fundamental building blocks. 
The SK basis of depth $d$ is the set of gate sequences up to length $d$. 
The SK basis for depth $d=3$ is shown in Figure~\ref{fig:tree huff v2}a. 
It should be noted that the null sequence \texttt{[]} is also a member of the basis, though it is not included in the table.
Revisiting our example of the unitary matrix from Equation~\ref{unitary matrix} and its decomposed circuit shown in Figure~\ref{fig:skd_circ}, the subsequent step involves generating the frequency distribution of gate sequences, derived from the representation of the quantum circuit in terms of gate sequences from the SK basis. 
The corresponding Huffman tree for the SK basis sequences is shown in Figure~\ref{fig:tree huff v2}b.
The tree is structured to minimize the path lengths for the most frequent sequences, thereby reducing the total number of bits required for the entire circuit description.
The Huffman codes for the SK basis elements are listed in Figure~\ref{fig:tree huff v2}c after traversal through the tree.

The number of bits required to describe the quantum circuit using Huffman coding of Solovay-Kitaev basis (v2) is $44$. 
This is again a substantial improvement over binary encoding, which requires $80$ bits, and over Huffman v1, which requires $59$ bits. 
The compression factor for this particular decomposition is $44/80 = 0.55$.


\begin{figure}[htb]
    \centering
    \includegraphics[width=\textwidth, clip, trim={0.5cm 0cm 0.2cm 7cm}]{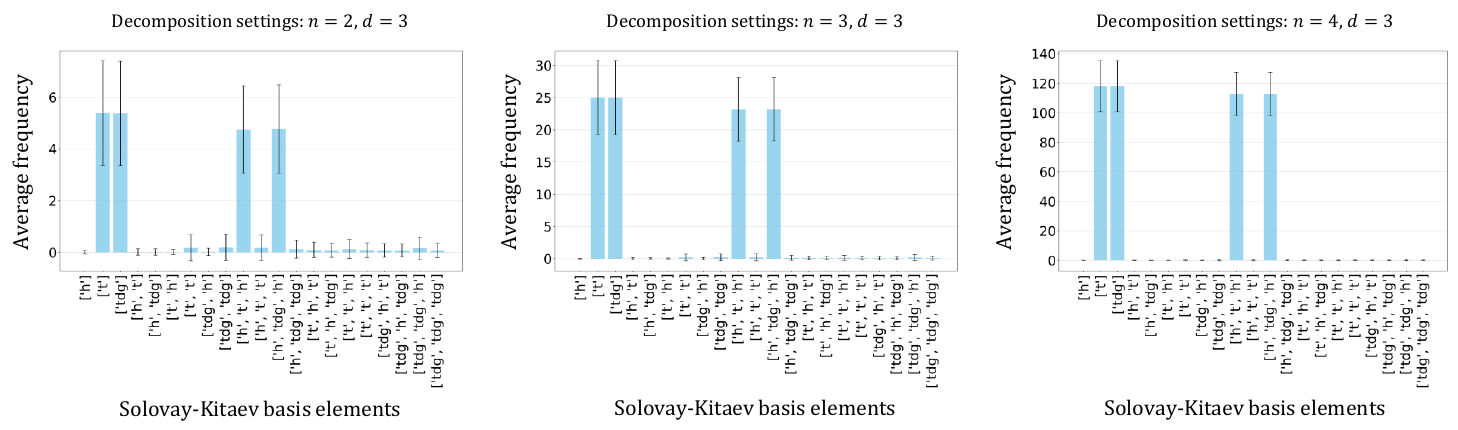}
    \caption{Bar charts showing the average distribution of frequencies of instructions from the Solovay-Kitaev basis in decomposed circuits obtained from Solovay-Kitaev decomposition of the data set of unitaries with $d=3$ and $n=2, 3, 4$.}
    \label{fig:avg-sk-basis}
\end{figure}

Analogous to the experiments to evaluate the generality of the generated Huffman codes in v1, we proceed with a similar analysis for v2 and perform the Solovay-Kitaev decomposition of the same data set of unitaries with depth $d = 3$ and degree of recursion $n=2, 3, 4$, and store the frequencies of usage for the instructions from the Solovay-Kitaev basis, as shown in Figure~\ref{fig:avg-sk-basis}.
Again, we observe consistency between the instructions most frequently used. 
This pattern of usage holds for a range of recursion depths $n$; however, it remains specific to each SK basis depth. 
For different depths, we would obtain a basis set different from the one in Figure~\ref{fig:tree huff v2}a, and consequently, a different distribution of usage frequencies. 
A significant remark on the observed average usage frequencies in Fig. \ref{fig:avg-sk-basis} is that there is only a handful of instructions that are used frequently compared to the remaining instructions in the set. 
This trend opens the possibility of omitting some basis-set elements during coding, thereby reducing the dictionary size. 
This motivates the final version of the EQISA encoding.

\subsubsection{Huffman encoded SK basis with frequency cutoff (v3)}
\label{huff_v3}

The third version of Huffman coding (v3) encodes the quantum circuit as a sequence of instructions in the Solovay-Kitaev basis, as in v2. 
However, the fundamental novelty of this version is the selection of instructions from the basis prior to Huffman encoding, with only the selected instructions being encoded.

The original gate set, [$H$, $T$, $T^\dagger$], is trivially included in this selection. 
Instructions appearing in the decomposed quantum circuit that do not belong to the selection are decomposed into the original basis gates, and their usage frequencies are updated. 

\begin{figure}[!bht]
    \centering
    \includegraphics[width = 0.65\textwidth,trim={8cm 0cm 0 2cm},clip]{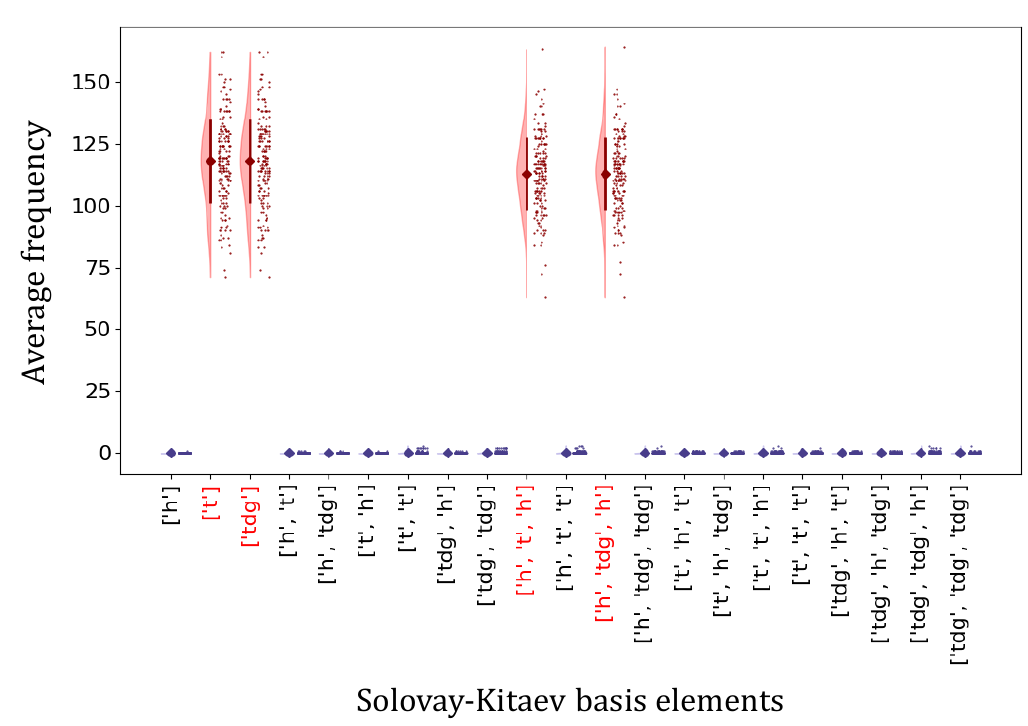}
    \caption{\centering Raincloud plot of usage frequencies of SK basis depth $d = 3$ instructions in decomposed circuits performed with $n = 4$. Instructions selected based on high frequency are marked in red.}
    \label{fig:raincloud}
\end{figure}

The usage frequencies of the SK basis of depth $d = 3$ are plotted in Figure~\ref{fig:raincloud}.
The selected Huffman instructions based on their frequencies are depicted in red.
This selection of instructions is made from the observed trend of usage frequencies over changing degrees of recursion presented in Fig. \ref{fig:avg-sk-basis}. 
The width of the rain cloud and the density of the scattered points represent the number of samples at a certain usage frequency. 
The mean and standard deviation of usage frequencies are depicted by the diamond markers and error bars, respectively. 
Note that the basis gates are necessarily included for encoding the pruned basis elements.
$H$ as an instruction has a low average usage, while the $H$ gate has a higher average frequency in the decomposed quantum circuits \ref{fig:avg-gateset-lists}. 
It can then be inferred that the dominant usage of the $H$ gate comes from the $HTH$ and $HT^\dagger H$ sequences in the SK basis.

The Huffman tree and table of codes for the selected Solovay-Kitaev basis instructions are presented in Figure~\ref{fig:huff v3}. 
The number of bits required to describe the quantum circuit using this version of Huffman coding is $45$, and the compression factor is $45/80 = 0.5625$. 
This number is only marginally higher than the number of bits required in v2 of Huffman coding ($44$), but the code dictionary is appreciably smaller, leading to faster decoding. 
In terms of generality, we observe a similar pattern of average frequency over different degrees of recursion $n$ for a set depth $d$. 
Therefore, a particular selection of instructions for the encoding also remains consistent for a particular depth $d$ over any choice of $n$. 

\begin{figure}[htb]
    \centering
    \begin{minipage}[t]{.45\textwidth}
    \vspace{0pt}
    \centering
        \begin{forest}
        for tree={
            draw,
            minimum size=1.75em,
            s sep=5mm
          },
          inner/.style={circle},  
          leaf/.style={rectangle, minimum width=2em, fill=gray!30},  
          EL/.style={edge label={node[midway,left,font=\scriptsize]{0}}},
          ER/.style={edge label={node[midway,right,font=\scriptsize]{1}}} 
          [20, inner
            [10, inner, EL
                [$HTH$, leaf, EL]
                [$T$, leaf, ER]
                ]
            [10, inner, ER
                [5, inner, EL
                    [$H$, leaf, EL]
                    [$T^\dagger$, leaf, ER]
                ]
                [$HT^\dagger H$, leaf, ER]
            ]
            ]
        \end{forest}
    \end{minipage}
    \hfill
    \begin{minipage}[t]{.45\textwidth}
    \vspace{0pt}
    \centering
        \renewcommand{\arraystretch}{1.0}
        \begin{tabular}{r|r}
            Selected basis & Huffman Code \\
            \hline
             $T$            &                    01 \\
             $HTH$          &                    00 \\
             $HT^\dagger H$ &                    11 \\
             $T^\dagger$    &                   101 \\
             $H$            &                   110  
        \end{tabular}
    \end{minipage}
    
    \begin{subfigure}[t]{.45\textwidth}
    \caption{Huffman tree representation for encoding selected quantum instructions.}
    \end{subfigure}
    \hfill
    \begin{subfigure}[t]{.45\textwidth}
    \caption{Huffman code table corresponding to adjoining tree.}
    \end{subfigure}
    
    \caption{\centering Huffman encoded SK basis with frequency cutoff (v3) for the example circuit.}
    \label{fig:huff v3}
\end{figure}

\subsubsection{Encoding for multi-qubit system}
\label{multiqubit}

Scaling up the decomposition and encoding routine to multi-qubit systems requires the stream of instructions describing the quantum circuit to also encode the qubit IDs. 
In this work, a simple binary encoding is chosen for encoding the qubit IDs with the number of bits for encoding scaling at $\lceil\log_2(N)\rceil$, where $N$ is the number of qubits in the quantum processor. 
The qubit ID stream is tailored according to the version of Huffman encoding adopted for the instruction stream, such that each opcode corresponds to one entry (or two in the case of $CX$) in the qubit ID stream. 
When scaling up this routine to much larger systems, Huffman coding can also be used to encode the qubit ID stream.
This is currently left as future work.

\subsection{Benchmarking results}
\label{sec_4.3}

So far, we have implemented and tested our decomposition and encoding routine on Haar-random unitaries. 
These circuits, though very general, do not necessarily have practical applications in the real world. 
In this section, we evaluate the performance of the decomposition and encoding routine for benchmark circuits from MQT Bench \cite{quetschlich2023mqtbench}, a curated library within the Munich Quantum Toolkit. 
The selection of benchmarks for our experiment consists of $79$ scalable benchmark circuits spanning system sizes from $2$ to $6$ qubits. 
The chosen benchmark circuits are listed in Appendix~\ref{appendix:bench}. 
These algorithms are developed for real-world use cases. 
We aim to study whether the algorithmic insights drawn from random circuits also apply to benchmark circuits of the same size and to explore whether we gain novel insights. 

\begin{figure}[htb]
    \centering
    \includegraphics[width=\textwidth, clip, trim={2cm 2.4cm 2cm 4cm}]{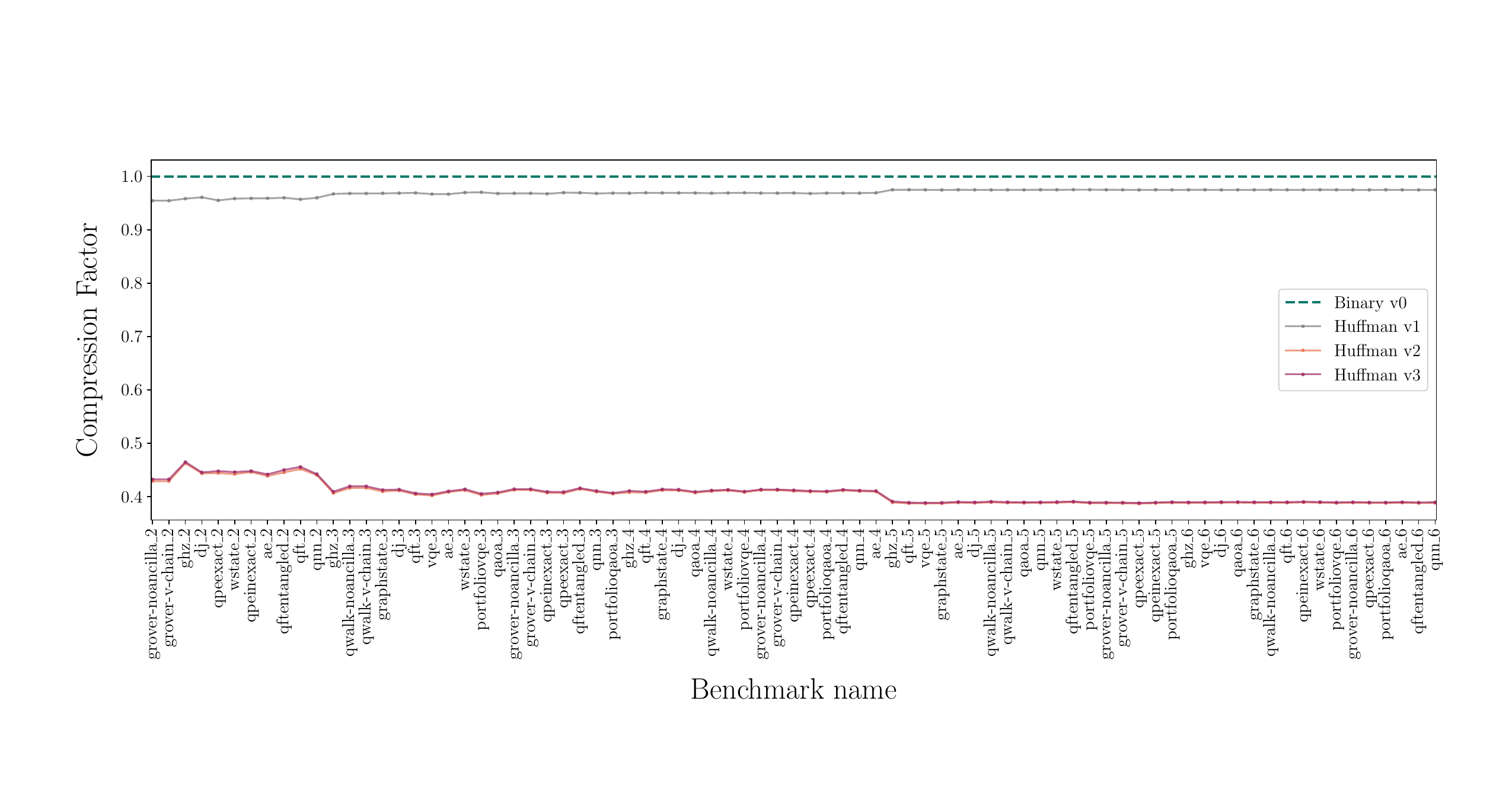}
    \caption{\centering Compression factor = $\frac{\texttt{len(huff\_encoded)}}{\texttt{len(bin\_encoded)}}$ for benchmark circuits from MQT Bench~\cite{quetschlich2023mqtbench}. Benchmarks sorted according to increasing system size.}
    \label{fig:cr_bench}
\end{figure}

We evaluated the performance of the Quantum Shannon Decomposition (for $n$-qubit decomposition in terms of CX and arbitrary rotation gates) and Solovay-Kitaev Decomposition (for $1$-qubit rotation decomposition to a discrete gate set) and selected the optimal parameters of depth $d = 5$ and degree of recursion $n = 4$. 
Note that the arguments and novelty presented apply to any choice of the parameters.
In Figure~\ref{fig:cr_bench}, the compression factor for each benchmark is compared across the three versions of Huffman coding, along with the baseline binary encoding.
The compression factor trends remain consistent across benchmarks of the same size, though there is greater variation in compression factor for smaller benchmarks than for larger systems. 
Note that the number of bits required to binary-encode the qubit IDs increases by $1$ when the system size increases from $2$ to $3$ and from $4$ to $5$.
However, for the v2 and v3 versions, a slight drop in the average compression factor is observed, as the increase in the qubit ID description is compensated by improved Huffman coding performance in larger systems. 
More importantly, the plot shows that Huffman v2 and v3 maintain a stable, low compression factor across all evaluated benchmarks, suggesting the robustness and effectiveness of our proposal across real-world circuit configurations. 

Another takeaway from these results is the consistency in the EQISA's performance on benchmarks compared to Haar-random unitaries.
The dataset size for real-world quantum algorithms is limited, whereas Haar-random unitaries can be instantiated efficiently and scalably.
Hence, Haar-random unitaries can be used to fix the encoding before applying the scheme and evaluating its performance on pragmatic algorithms.

\subsection{Lossless compression}
\label{bzip}

Lossless compression ensures no loss of information during compression, allowing the original data to be exactly reconstructed upon decompression. 
These algorithms operate by searching for patterns and redundancies in data that can be expressed more concisely. 
bzip2 \cite{seward1996bzip2}, developed by Julian Seward in 1996, is an efficient and popular lossless compression technique attributed to its excellent balance of compression efficiency and speed. 
The bzip2 pipeline consists of a sequence of transformations that rearrange and encode the data. 
These steps and their order are designed to complement one another and improve overall efficiency. 
The distinguishing innovation of bzip2 is the Burrows-Wheeler transform~(BWT)~\cite{burrows1994block} that rearranges the data to have runs of similar symbols and improves the effectiveness of the subsequent move-to-front transform~(MTF) and run-length encoding~(RLE). 
As a result, bzip2 is very efficient at compressing text that contains repeated patterns and is asymmetric; decompression is faster than compression. 
These qualities make it a suitable choice for compressing QASM instructions in quantum circuits, which contain repeated instances of gate operations and qubit IDs. 

After achieving significant data compression with various Huffman encoding techniques tailored to quantum circuits, we apply further compression with bzip2. 
Given the nature of encoded quantum instruction streams, which often contain repetitive sequences of 0s and 1s, bzip2 can further optimize the transmission efficiency of these streams.
The process involves using the selected encoding (binary or one of the three Huffman variants described earlier) and compressing it with bzip2.

\begin{figure}[hbt]
    \centering
    \includegraphics[width=1\textwidth, clip, trim={0cm 0cm 0cm 0cm}]{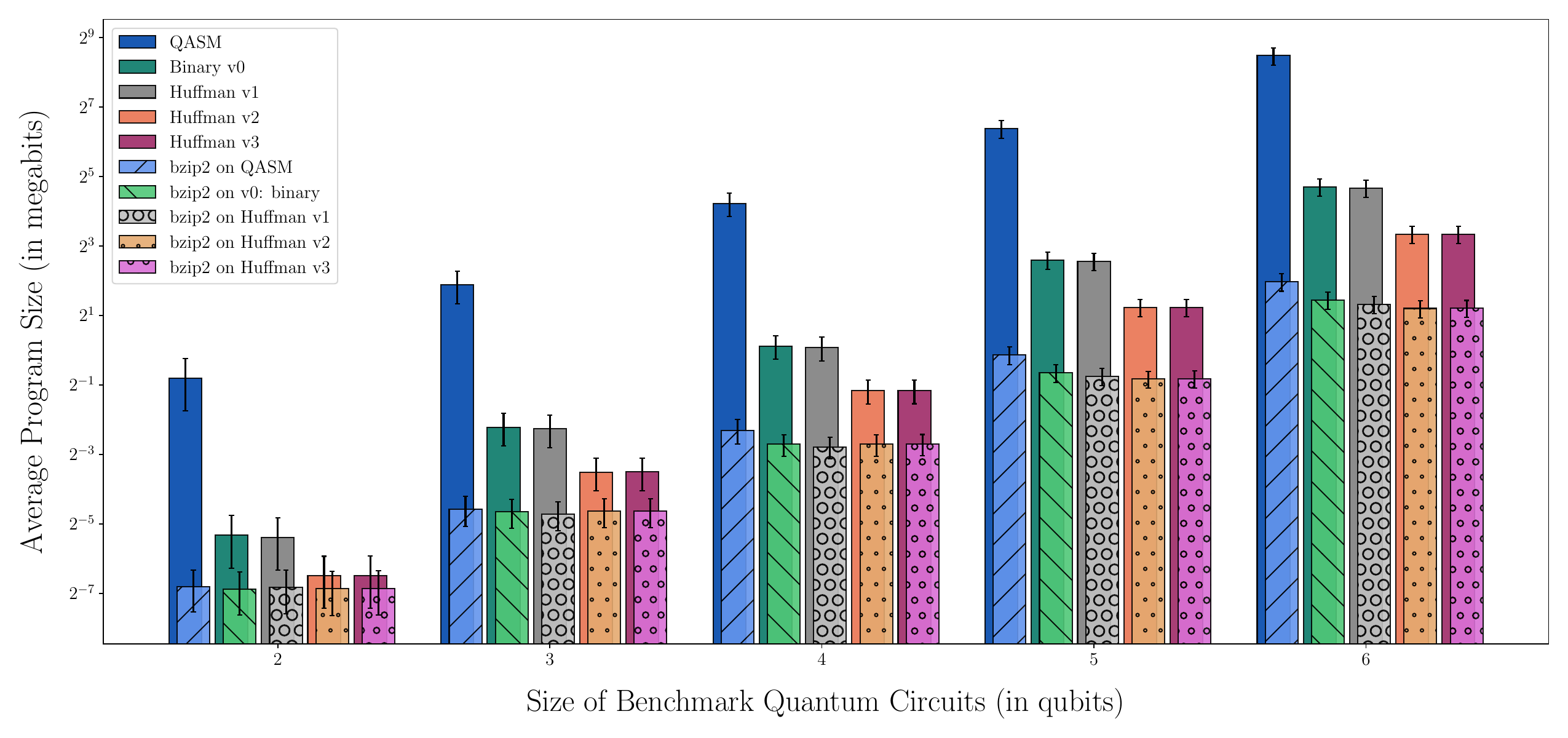}
    \caption{Average program sizes for benchmark circuits from MQT Bench~\cite{quetschlich2023mqtbench} grouped according to the size of the system in qubits. Overlaid bars depict compressed program sizes after bzip2.}
    \label{fig:program_sizes}
\end{figure}

The average original and the bzip2-compressed program sizes are compared on a $\log_2$ scale in Figure~\ref{fig:program_sizes} to assess the effectiveness of applying a secondary compression stage. 
The additional compression achieved suggests that layering compression techniques, starting with Huffman encoding tailored to capture the quantum `context' of circuits effectively, followed by bzip2, provides a robust method for minimizing quantum program size. 
As a summary, Huffman v3 with bzip2 is the proposed effective EQISA.

The methods and benchmarks presented in this section set the stage for evaluating the proposed EQISA based on Huffman v3 on practical quantum computing tasks in the following section.

\section{Applications of EQISA}
\label{applications}

In this section we motivate the applications of the proposed EQISA from three pragmatic perspectives.
We discuss its impact on energy in quantum cryogenic control, in the discovery of contextual abstractions in quantum circuits, and in the estimation of information-theoretic measures for quantum computation.
 
\subsection{Energy gains in cryogenic control}
\label{application1_energy_gains}

Quantum control architecture is a crucial component~\cite{anastasia} in the development of a full-stack quantum computer. 
Control is performed using various methods: superconducting qubits employ microwave pulses, trapped ions are manipulated with laser pulses, photonic qubits are controlled via optical devices such as beam splitters, and silicon quantum dots are manipulated through electrical pulses. 
Most of the systems, typically CMOS circuits and FPGAs, operate at extremely low temperatures close to absolute 0 Kelvin~\cite{cryo-cmos,Xue2021,cryo_fpga}. 
One of the most important limitations of these devices is the tight power-consumption budget associated with their cryogenic operating temperatures. 
Employing these devices at temperatures far below their rated operating range is challenging due to anomalous behavior and non-idealities in their I-V characteristics~\cite{cryo_fpga}. 
Dilution refrigerators, which are commonly used to cool quantum processors to sub-Kelvin temperatures, are highly sensitive to device power consumption. 
At these temperatures, even small amounts of heat can significantly impact the system's performance, including introducing thermal noise and reducing the coherence times of quantum states. 
The power dissipation budget of dilution refrigerators for cooling quantum processors determines the maximum allowable heat load to maintain ultra-low temperatures. 
It ranges from around 1W at 4K and goes down to the order of several $\mu$W at 10 mK~\cite{cryoCMOS_Hart}.
This energy bottleneck of cryogenic control necessitates the development of efficient methods for transmitting instructions from room temperature (RT) (around 300K) to cryogenic environments (typically around 4K). 
Different technologies have been proposed in recent years for establishing high-speed, low-power transmission links between RT and 4K. 
The energy per bit of data ranges from a few hundred fJ/bit~\cite{optic_datarate1, electro-optic_modulation} using optical fibers with photonic links and CMOS-based transceiver chip~\cite{10067445} to a few pJ/bit using CMOS DAC-based wireline transmitter~\cite{cmos_wireline}.

QISA is analogous to classical instruction set architectures, such as Intel x86, ARM, and RISC-V, and serves as a crucial intermediary between higher-level quantum programming languages and lower-level quantum hardware operations. 
It provides a more comprehensive level of abstraction for describing the physical details of the microarchitecture and hardware than high-level quantum programs and algorithms do.
Notable contributions to this field have been made in \cite{anastasia}, where the authors propose a scalar quantum extension termed QUASAR and a vector extension qV based on the existing RISC-V ISA~\cite{risc-manual}. 
The efficacy of their proposed QISA is measured by metrics such as encoding efficiency and execution time. 
Assessment of ISAs by characterizing the control processor's performance in terms of parameters such as circuit complexity, gate density, and diversity provides an algorithmic evaluation of the QISA. 

Our proposed Huffman-encoded quantum instruction sets, followed by bzip2 compression, offer a redefined, dense (along the depth) representation of quantum circuits in terms of Solovay-Kitaev basis instructions. 
The reduced bit-level throughput requirement for the encoded instructions provides a key advantage in transmitting the instruction stream from room temperature to the control architecture's operating temperature.
Here, we estimate the energy gains for the cryo-CMOS DAC-based wireline transmitter technology presented in \cite{cmos_wireline}, which achieves an energy rate of 2.46 pJ/bit and a data rate of 40 Gbps.

\begin{figure}[htb]
    \centering
    \includegraphics[width=0.5\textwidth, clip, trim={1cm 0cm 1cm 1.5cm}]{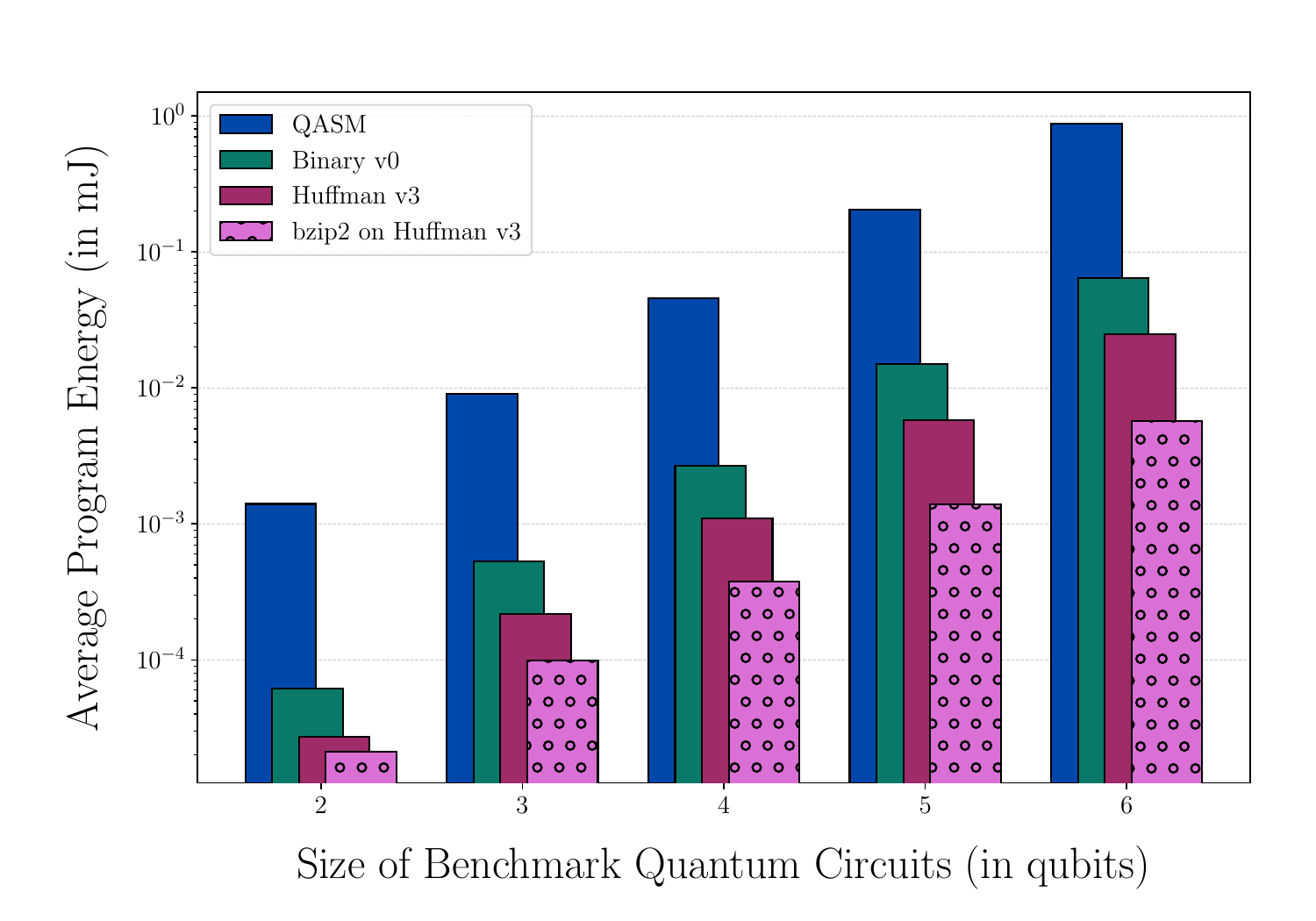}
    \caption{\centering Average energy consumption for MQT Bench~\cite{quetschlich2023mqtbench} quantum program transmission represented as: QASM, Binary v0 encoded, Huffman v3 encoded, and bzip2 compression on Huffman v3 encoded data. Each bar represents the average energy consumption for a given system size in qubits, with bars for each format overlaid for direct comparison. Energy per bit is taken to be 2.46 pJ/bit~\cite{cmos_wireline}.}
    \label{fig:program_energy}
\end{figure}

Figure \ref{fig:program_energy} depicts the average energy load (in log-scale) for transmission of the quantum programs (instruction stream plus qubit ID stream) from MQT Bench~\cite{quetschlich2023mqtbench}. 
Adopting the proposed v3 Huffman coding brings down the heat load to approximately 2.5 \% of that incurred by QASM format (and 40\% of that incurred by binary encoded format).
Performing lossless compression (bzip2) on the Huffman v3 encoded instructions brings the heat load down to around 1 \% of that by the QASM. 
Paying attention to some caveats, the subsequent bzip2 compression adds additional overhead for decompression at the cryo level. 
bzip2 is asymmetric, taking less time to decompress than to compress, which makes it a good choice for lossless compression. 
Huffman v3, which offers asymptotically identical compression to v2, is a better choice for encoding because of its shorter code dictionary. 

The diminished heat load during data transmission signifies a promising avenue for scalability in program size. 
As we envisage the era of fault-tolerant quantum computation, the complexity and size of quantum programs are poised to expand significantly, encompassing logical-level operations. 
The reduced energy overhead from instruction transmission alleviates stringent power constraints in cryogenic environments, particularly on control processors.

\subsection{Discovering high-level circuit abstractions}
\label{application2_abstractions}

Studying high-level quantum programming abstractions is crucial for advancing the field of quantum computing. 
This is akin to understanding language through common substrings or patterns, which simplifies the creation and comprehension of new words or phrases.
As an analogy, if we observe the sequence of letters h-o-o-d occurring in many words, such as fatherhood, motherhood, and brotherhood, we ground the meaning of ``the state of being'' and thereby create and use new words like adulthood, childhood, and likelihood. 

In classical programming, identifying common patterns, such as loops and conditional statements, has led to the development of higher-level programming constructs, like for-each loops and lambda expressions, that simplify coding tasks and enhance readability and maintainability.
Similarly, in quantum computing, recognizing and formalizing frequently occurring sequences of quantum gates (like those from the Solovay-Kitaev (SK) basis) as high-level constructs can provide similar benefits. 
These sequences, once identified and abstracted, can serve as fundamental building blocks or macros, which encapsulate complex quantum operations into single, reusable components. 

\begin{table}[htb]
\centering
\renewcommand{\arraystretch}{1.1}
\scriptsize
\begin{tabular}{|l|l|l|}
\hline
Selection for $d=4$ & Selection for $d=5$ & Selection for $d=6$ \\
\hline
    $CX$ & $CX$ & $CX$ \\ 
    $H$ & $H$ & $H$ \\
    $T$ & $T$ & $T$ \\
    $T^\dagger$ & $T^\dagger$ & $T^\dagger$ \\
    $HTH$ & $HTH$ & $HTH$ \\
    $HT^\dagger H$ & $HT^\dagger H$ & $HT^\dagger H$ \\
    $HTHT$ & $HTHT$ & $HTHT$ \\
    $HTHT^\dagger$ & $HTHT^\dagger$ & $HTHT^\dagger$ \\
    $HT^\dagger HT$ & $HT^\dagger HT$ & $HT^\dagger HT$ \\
    $HT^\dagger HT^\dagger$ & $HT^\dagger HT^\dagger$ & $HT^\dagger HT^\dagger$ \\
    $THTH$ & $THTH$ & $THTH$ \\
    $THT^\dagger H$ & $THT^\dagger H$ & $THT^\dagger H$ \\
    $T^\dagger HTH$ & $T^\dagger HTH$ & $T^\dagger HTH$ \\
    $T^\dagger HT^\dagger H$ & $T^\dagger HT^\dagger H$ & $T^\dagger HT^\dagger H$ \\
    & $THTHT^\dagger$ & $THTHT^\dagger$ \\
    & $THT^\dagger HT^\dagger$ & $THT^\dagger HT^\dagger$ \\
    & $T^\dagger HTHT$ & $T^\dagger HTHT$ \\
    & $T^\dagger HT^\dagger HT$ & $T^\dagger HT^\dagger HT$ \\
    &  & $THTHT^\dagger T^\dagger$ \\
    &  & $THT^\dagger HT^\dagger T^\dagger$ \\
    &  & $TTHTHT^\dagger$ \\
    &  & $TTHT^\dagger HT^\dagger$ \\
    &  & $T^\dagger HTHTT$ \\
    &  & $T^\dagger HT^\dagger HTT$ \\
    &  & $T^\dagger T^\dagger HTHT$ \\
    &  & $T^\dagger T^\dagger HT^\dagger HT$ \\
\hline
\end{tabular} 
\caption{Selected SK basis for encoding for varying system sizes. Note that larger dictionaries expand on smaller ones.}
\label{selections}
\end{table}

Analyzing average frequencies of usage of SK basis instructions leads to the identification of instructions that are used much more frequently than others, as depicted in Fig. \ref{fig:raincloud} for the case of SK basis depth $d=4$. 
This observation served as the basis for v3 of Huffman encoding, in which we selected the most frequently occurring SK basis instructions and included only them in the code dictionary to describe the circuit. 
These selected instructions for SK basis of depth $d=4,5,6$ are depicted in Table~\ref{selections}. 
The instructions are selected by thresholding the average usage frequency across a dataset of $200$ random unitaries. 
We observe that the gate sequences at lower depths also have a consistently high usage at higher depths. 

Similar approaches to discovering programming abstractions through program synthesis have been explored in \cite{sarra2023discovering,kundu2026reinforcement}, where deep reinforcement learning is used to train the compiler to synthesize unitaries. 
The library of gates is constantly updated and used to solve similar synthesis problems.
Finding these abstractions not only streamlines the quantum programming process but also facilitates the development of new algorithms by reusing and combining these high-level constructs.
Such a workflow is particularly suited for automated quantum program synthesis, such as via quantum architecture search, where the semantics of the abstractions are not essential.
These abstractions can serve as a basis for training foundational models~\cite{apak2024ketgpt} to generate quantum algorithmic structures.

\subsection{Estimating total complexity of quantum algorithms}
\label{application3_uncomplexity}

\begin{figure}[htb]
    \centering
    \includegraphics[width=\textwidth, clip, trim={1cm 2.6cm 2cm 4cm}]{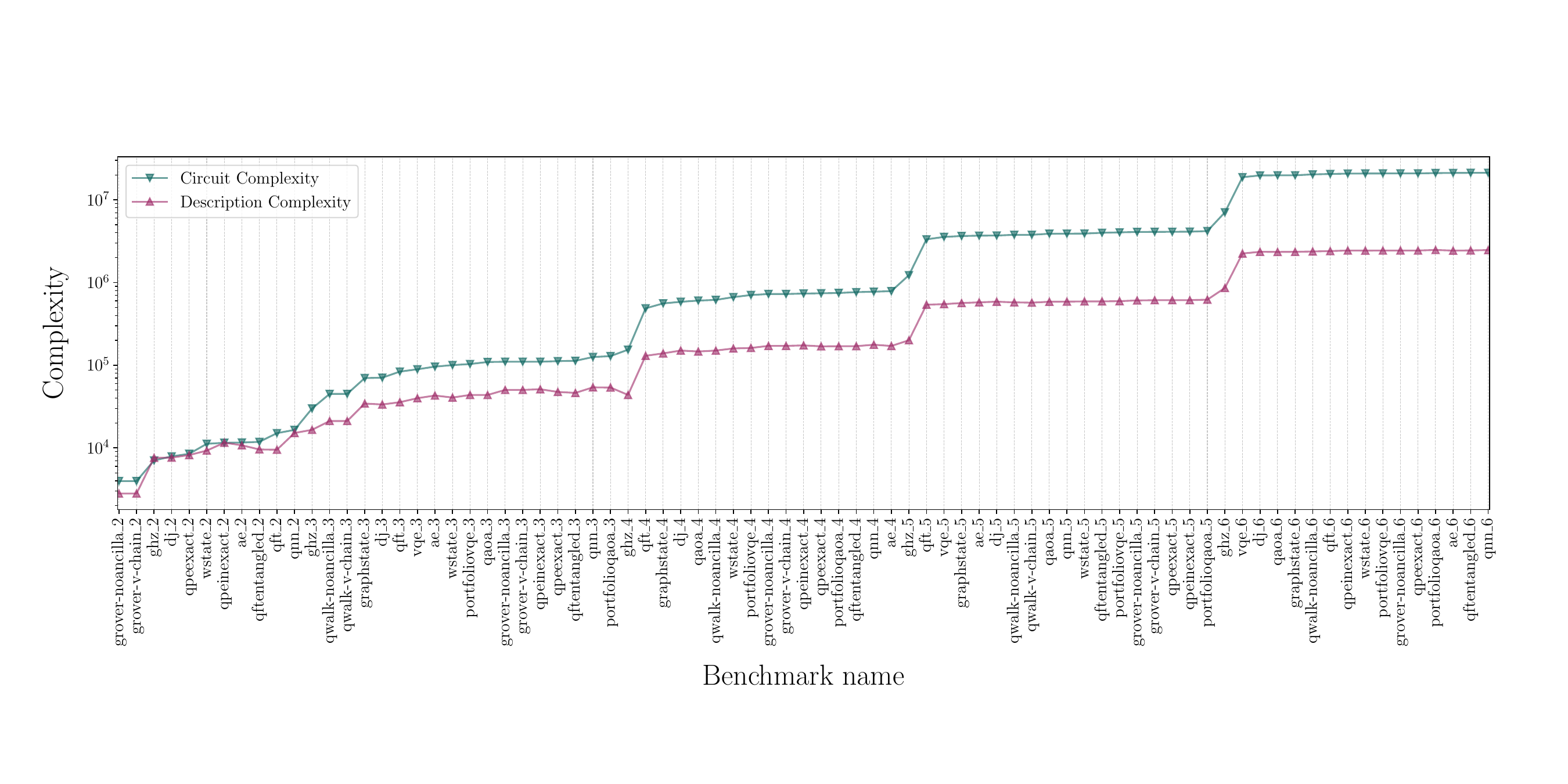}
    \caption{\centering Circuit complexity and description complexity plotted for benchmark circuits~\cite{quetschlich2023mqtbench}, arranged according to increasing qubit size of the benchmarks.}
    \label{fig:complexities_bench}
\end{figure}

From a theoretical perspective, the decomposition and encoding routine is an estimate for the quantum description complexity of the underlying quantum transformation. 
For a practical estimate of description complexity, it adheres to the criteria of minimality via Huffman coding, and to universality and invariance via the Solovay-Kitaev theorem.
Based on these criteria, the Huffman encoding of SK instructions provides a practical and effective method for approximating the description complexity of quantum circuits. 
Next, inspired by \cite{brown2018second}, we aim to describe the total quantum complexity as the sum of the quantum circuit complexity and quantum description complexity. 
Quantum circuit complexity is defined as the product of the system size (number of qubits) and the circuit depth, i.e., the quantum volume. 
Adopting this definition as the measure of circuit complexity and Huffman v3 as the measure for description complexity, the circuit and description complexity for benchmarks are presented in Figure~\ref{fig:complexities_bench}.

We observe that the description complexity (marked in purple) does not increase as rapidly as the circuit complexity (marked in cyan) with the growing system size. 
This suggests that the Huffman encoding of SK basis instructions successfully captures significant contextual or abstract information about quantum circuits, thereby simplifying their representation without losing essential details necessary for accurate computation.
It is insightful to find that the smaller the gap between these two complexities (within a specific qubit size), the more irregular or incompressible the circuit is.
For example, the qnn (quantum neural network) has less structure than ghz state preparation.

\section{Conclusion}
\label{conclusion}

Driven by both industrial and theoretical motivations, this article presents two major contributions: (i) an energy-efficient quantum instruction set architecture (EQISA), and (ii) an estimate of the description complexity of quantum circuits. 
The proposed EQISA enhances the compilation and control performance for optimal firmware design. 
We investigate how the representation of decomposed quantum circuits can be compressed using tools from algorithmic information theory.
For a specific target native gate set, EQISA represents the decomposed circuits of the quantum unitary transforms as Huffman-encoded opcodes in the corresponding Solovay-Kitaev basis. 
Three different versions of Huffman coding, with increasing levels of sophistication, are developed and presented, along with a comparative analysis of the average compression factor over datasets of random unitaries for various hyperparameters of the Solovay-Kitaev decomposition. 
The efficacy of the decomposition and encoding routine is then portrayed on a set of real benchmark circuits. 
We observe consistent compression performance across a range of benchmarks, which indicates the generality of these encoding schemes. 
A subsequent lossless compression with bzip2 over Huffman-encoded instruction streams further improves compression and reduces the total program size.

As an application, we highlight the crucial role of a cryogenic control architecture in enabling scalable, fully integrated quantum computation at qubit operation temperatures. 
EQISA addresses the energy bottlenecks and constraints within these systems, where the compressed quantum instruction streams reduce the heat load of instruction transmission to cryogenic levels. 
Additionally, the Huffman-encoded representation of decomposed circuits also provides a practical measure of quantum description complexity.
We observe that it successfully captures high-level quantum programming abstractions, offering valuable insights into the semantics of quantum algorithms and fostering the development of novel algorithms.

This project was conceived within the context of a broader initiative aimed at optimal quantum firmware design alongside explainable quantum circuits~\cite{xie2025deqompile}, gate set optimization~\cite{sarkar2024yaqq}, and energy-efficient pulse control~\cite{fauquenot2025open}.
Here, we discuss some promising future directions of this work.

\begin{itemize}[nolistsep,noitemsep]

    \item Though the decomposition and encoding framework is designed to work for any general gate set, the canonical gate set of \{$H$, $T$, $T^\dagger$\} has been chosen in this work for consistency with the current pipeline. 
    The YAQQ framework~\cite{sarkar2024yaqq}, which operates at a level just above EQISA in the optimal firmware suite, searches for an optimal gate set over an ensemble of random unitaries by minimizing a cost function that has the parameters fidelity, circuit depth, and novelty.
    Adopting gate sets based on YAQQ can further increase decomposition fidelity and reduce circuit depth, amplifying the benefits of EQISA.
    
    \item Designing the EQISA to be aware of the microarchitecture and control systems makes it more comprehensive~\cite{Chong2017} in describing the program's physical details. 
    An extension of this project could integrate the EQISA framework with a microarchitecture to translate the compressed, encoded representation of decomposed quantum circuits to the pulse level.
    In this aspect, EQISA can be integrated with the EO-GRAPE and EO-DRLPE methods~\cite{fauquenot2025open} for energy-efficient pulse control as a layer below, delivering optimized pulse instructions for the compressed circuit representations. 

    \item The concept of description uncomplexity as a resource for performing computation is put forth in \cite{brown2018second}.
    Following these ideas, the algorithmic randomness for quantum circuits can be estimated as proportional to the incompressibility. 
    Quantum circuits with low algorithmic randomness can be inferred to have a greater untapped resource for performing the target computation. 
    This analogy can be used to empirically assess the potential for quantum supremacy in circuits with high algorithmic randomness, as well as the expressivity of ansatzes in parametric quantum circuits.
    For n-qubit unitary matrices, while the general scaling of decomposition to discrete and local gates scales exponentially, it is imperative that specifying such quantum computation (either as a unitary matrix or a sequence of gates) is intractable.
    Thus, the EQISA can be constructed from a restricted, resource-bounded class of quantum circuits that can be pragmatically demonstrated on near-term quantum hardware.
\end{itemize}

\section*{Software availability} \label{software}

The open-sourced code for EQISA, configuration files, output data, and plotting codes for the experiments presented in this article are available at:
\href{https://github.com/Advanced-Research-Centre/EQISA}{https://github.com/Advanced-Research-Centre/EQISA}.

\section*{Acknowledgements}

The authors thank Fabio Sebastiano, Ramon Overwater and İlker Polat for insightful discussions regarding the cryogenic control energy bottleneck. 
A.S. acknowledges funding from the Dutch Research Council (NWO) through the project ``QuTech Part III Application-based research" (project no. 601.QT.001 Part III-C—NISQ). 

\section*{Author contributions}


Conceptualization, A.S.; 
Methodology, A.S. and S.M.; 
Software, S.M.; 
Formal Analysis, A.S. and S.M.; 
Investigation, A.S. and S.M.; 
Writing – Original Draft Preparation, S.M. and A.S.; 
Writing – Review \& Editing, A.S. and S.F.; 
Visualization, S.M.; 
Supervision, A.S. and S.F.; 
Funding Acquisition, S.F.

\bibliographystyle{unsrt}
\bibliography{ref.bib}

\appendix
\newpage
\section{Quantum Shannon decomposition}
\label{appendix:qsd}

The quantum Shannon decomposition~(QSD)~\cite{shende2005synthesis} is a technique for expressing any $n$-qubit quantum operator as an exact decomposition of single-qubit rotations and 2-qubit controlled gates. The algorithm follows a divide-and-conquer strategy in a recursive fashion and breaks down the $n$-qubit unitary matrix into smaller matrices. The algorithm starts with cosine-sine decomposition~(CSD), a well-known technique in linear algebra that divides the target matrix $U$ into smaller blocks. The algorithm recursively performs CSD and other decomposition techniques such as eigenvalue decomposition and Euler decomposition to eventually express the original complex operator as a sequence of single-qubit gates and $CX$ gates that can be passed as an executable stream of instructions for the hardware. This synthesis technique functions as a quantum version of the classical Shannon decomposition of Boolean functions. 
 
\begin{figure}[htb]
    \centering
    \captionsetup{justification=centering}
    \includegraphics[width=8.5cm]{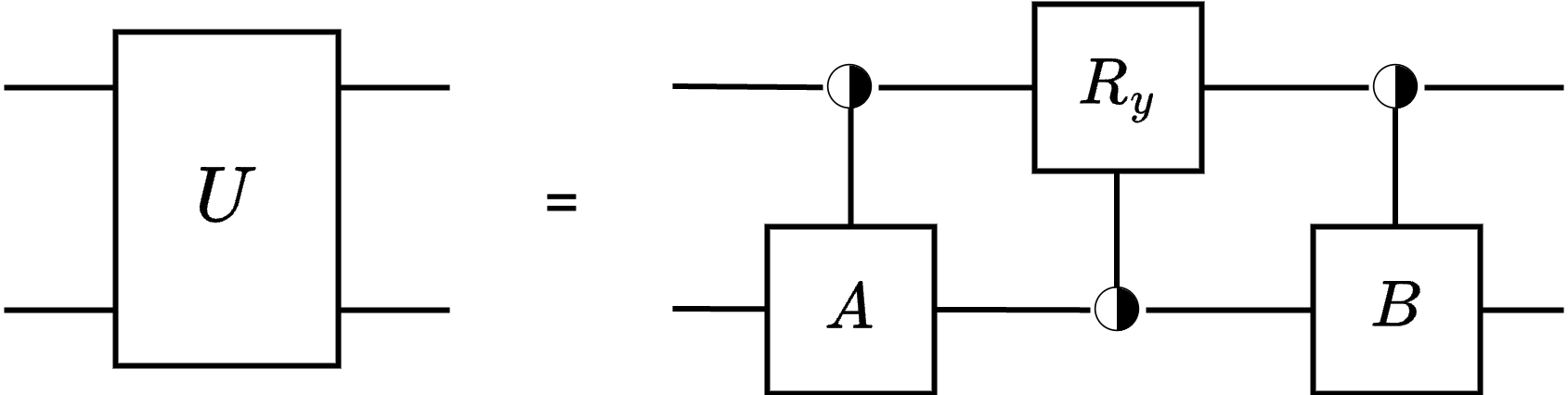}
    \caption{The Cosine-Sine decomposition acting on $n$-qubit gate $U$.}
    \label{cosine-sine decomposition}
\end{figure}

A quantum operation on $n$-qubits is represented by a unitary matrix $U$ of dimensions $2^n\times2^n$. According to CSD, $U = LMR^\dagger$ where $L$ and $R$ are block-diagonal matrices representing uniformly controlled gates and the middle matrix $M$ that represents a controlled $R_y$ rotation on the most significant bit (MSB).  

\begin{equation}
    \label{csd}
    U = 
    \begin{bmatrix}
        \begin{array}{c|c}
        U_{00} & U_{01} \\ \hline
        U_{10} & U_{11}
        \end{array}
    \end{bmatrix}
    = 
    \begin{bmatrix}
        \begin{array}{c|c}
        L_1 & 0 \\ \hline
        0 & L_2
        \end{array}
    \end{bmatrix}
    \begin{bmatrix}
        \begin{array}{c|c}
        C & -S \\ \hline
        S & C
        \end{array}
    \end{bmatrix}
    \begin{bmatrix}
        \begin{array}{c|c}
        R_1 & 0 \\ \hline
        0 & R_2
        \end{array}
    \end{bmatrix}^\dagger
\end{equation}
According to Equation~\ref{csd}, $L_1$, $L_2$, $R_1$ and $R_2$ are unitary matrices of size $2^{n-1}$. $C$ and $S$ are diagonal matrices such that $C^2+S^2 = I$, thereby justifying the name of the decomposition technique. 
The matrices $L$ and $R$ are termed as quantum multiplexors and they enact $L_1$ ($R_1$) or $L_2$ ($R_2$) conditioned on the state of the MSB. The middle matrix resembles the $R_y$ rotation matrix that is targeted on the MSB and controlled by the states of the lower-order qubits.
In terms of quantum circuit, the CSD can be decomposed as shown in Figure~\ref{cosine-sine decomposition}, with controlled-$A$ representing $R\dagger$ and controlled-$B$ representing $L$.

\begin{figure}[htb]
    \centering
    \captionsetup{justification=centering}
    \includegraphics[width=5.5cm]{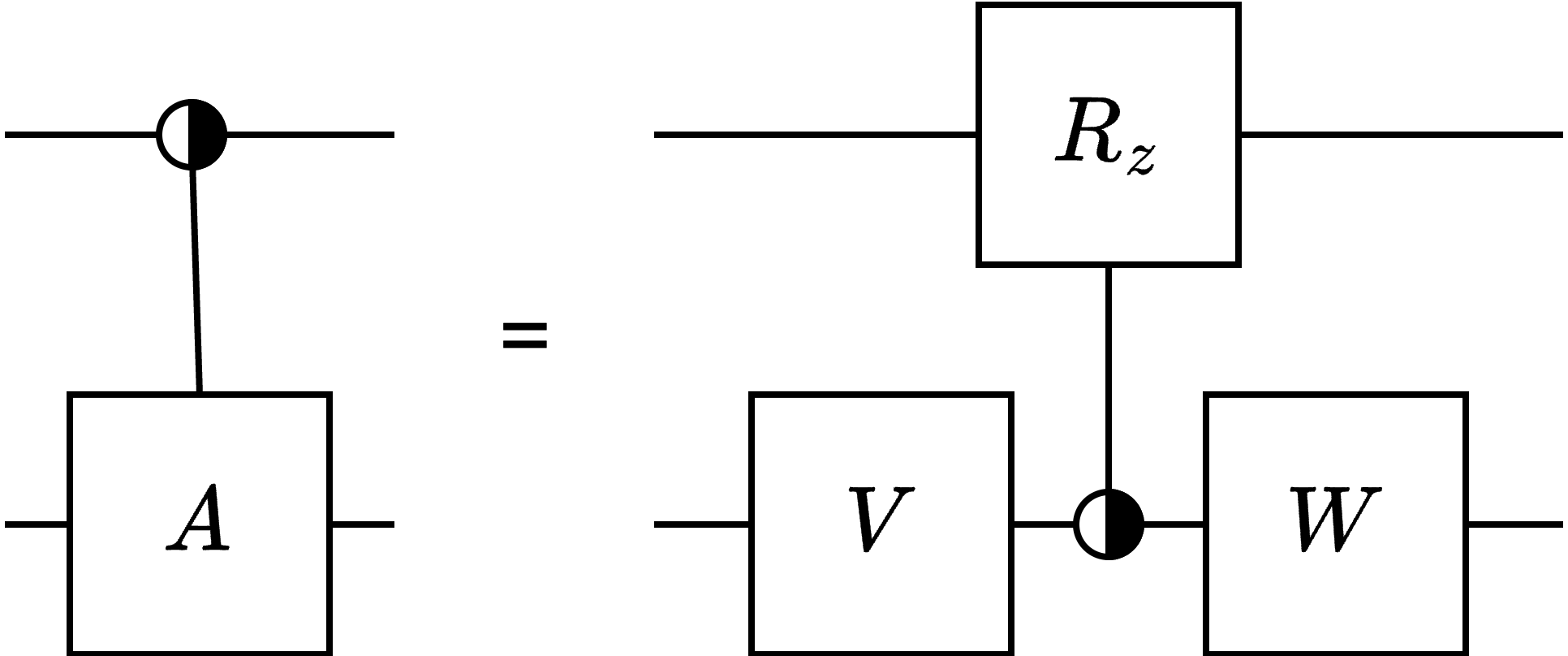}
    \caption{Demultiplexing of a multiplexor.}
    \label{qsd_demultiplex}
\end{figure}

The left and right gates (controlled-$A$ and controlled-$B$) further undergo a demultiplexing routine that performs an eigenvalue decomposition of the matrices as,
\begin{equation}
    \label{demux}
    \begin{bmatrix}
        A_1 & 0 \\ 
        0 & A_2
    \end{bmatrix}
    = 
    \begin{bmatrix}
        P & 0 \\ 
        0 & P
    \end{bmatrix}
    \begin{bmatrix}
        \Lambda & 0 \\ 
        0 & \Lambda^\dagger
    \end{bmatrix}
    \begin{bmatrix}
        Q & 0 \\ 
        0 & Q
    \end{bmatrix}
\end{equation}
where $P$ and $Q$ are unitary matrices, and $\Lambda$ is a unitary diagonal matrix. The left and right matrices ($P$ and $Q$) can be represented as quantum gates operating on the lower-order qubits and independent of the MSB. The middle matrix corresponds to a $R_z$ operation on the MSB controlled by the lower qubits.
This is shown in Figure~\ref{qsd_demultiplex}.

This process of CSD, followed by subsequent demultiplexing operation, is performed recursively until the algorithm reaches the base case. At the base level, the operator sequence consists of only single qubit gates. At this point, any single-qubit unitary operation is changed into a rotation gate following Euler decomposition. 
The implementation of the algorithm is presented with two optimization strategies. Following the first strategy, the multiplexed $R_y$ operation in Figure~\ref{cosine-sine decomposition} is implemented using $CZ$ gates. In the second strategy, the recursion is stopped at the level of 2-qubit operations, and the resulting circuit is decomposed into $CX$ gates and single-qubit rotation gates. With the optimization strategies, the algorithm yields an efficient synthesis of complicated quantum operators with a minimized number of $CX$ gates.

\begin{figure}[htb]
    \centering
    \captionsetup{justification=centering}
    \includegraphics[width=12cm]{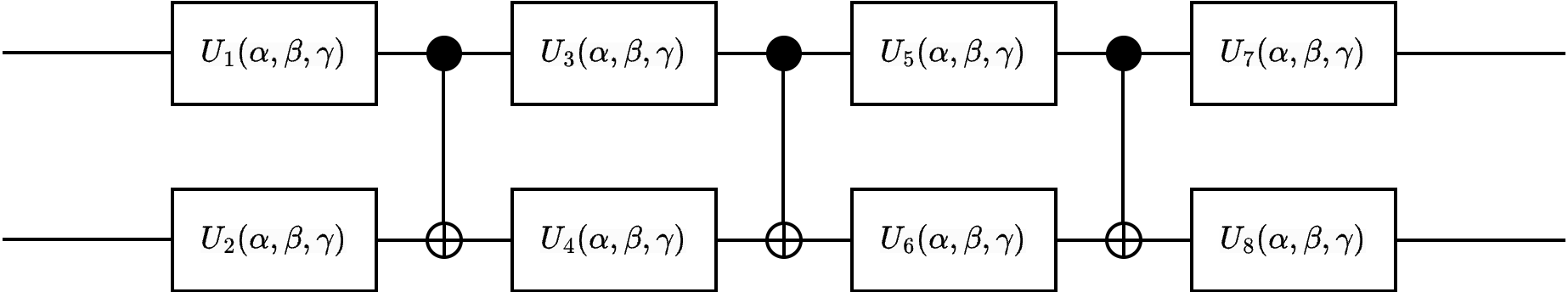}
    \caption{Final decomposed form of a 2-qubit arbitrary unitary operator expressed as a sequence of 1-qubit rotations and $CX$ gates.}
    \label{2q_QSD}
\end{figure}

In this project, the Qiskit implementation of the QSD algorithm is employed. The QSD process is fixed to run with the second optimization as mentioned above, and it returns a decomposed quantum circuit consisting of single qubit unitary rotations and $CX$ gates as depicted in Figure~\ref{2q_QSD} for a 2-qubit system. At this point, we perform Solovay-Kitaev decomposition (SKD) on the single-qubit unitary rotations to express them in terms of the native gates \{$H$, $T$, $T^\dagger$\}, and leave the $CX$ gates untouched. As a result, the multi-qubit random unitary operator is expressed as a quantum circuit built of gates from the discrete gate set \{$H$, $T$, $T^\dagger$, $CX$\}. Adopting SKD for breaking down the intermediate gates enables the application of the modified alphabet of instructions from the SK basis to express the decomposed quantum circuit.

\section{Benchmark circuits}
\label{appendix:bench}

List of benchmark circuits taken from the MQT Bench~\cite{quetschlich2023mqtbench} library. 
The suffix \texttt{\_indep\_qiskit} denotes that the circuits are described at a target-independent level using Qiskit as the compiler. 
The chosen benchmarks are all scalable algorithms with system sizes ranging from 2 to 6 qubits.

\begin{table}[htb]
\small
\renewcommand{\arraystretch}{1.0}
\begin{tabular}{||l|l||}
\hline
\hline
Benchmark Codename & Circuit Name and Size \\
\hline
\hline
    \texttt{ae\_indep\_qiskit\_2} & Amplitude Estimation, 2 qubit \\ 
    \texttt{ae\_indep\_qiskit\_3} & Amplitude Estimation, 3 qubit \\ 
    \texttt{ae\_indep\_qiskit\_4} & Amplitude Estimation, 4 qubit \\ 
    \texttt{ae\_indep\_qiskit\_5} & Amplitude Estimation, 5 qubit \\ 
    \texttt{ae\_indep\_qiskit\_6} & Amplitude Estimation, 6 qubit \\ 
\hline
    \texttt{dj\_indep\_qiskit\_2} & Deutsch Josza, 2 qubit \\ 
    \texttt{dj\_indep\_qiskit\_3} & Deutsch Josza, 3 qubit \\ 
    \texttt{dj\_indep\_qiskit\_4} & Deutsch Josza, 4 qubit \\ 
    \texttt{dj\_indep\_qiskit\_5} & Deutsch Josza, 5 qubit \\ 
    \texttt{dj\_indep\_qiskit\_6} & Deutsch Josza, 6 qubit \\
\hline
    \texttt{ghz\_indep\_qiskit\_2} & GHZ State, 2 qubit \\
    \texttt{ghz\_indep\_qiskit\_3} & GHZ State, 3 qubit \\    
    \texttt{ghz\_indep\_qiskit\_4} & GHZ State, 4 qubit \\    
    \texttt{ghz\_indep\_qiskit\_5} & GHZ State, 5 qubit \\    
    \texttt{ghz\_indep\_qiskit\_6} & GHZ State, 6 qubit \\
\hline
    \texttt{graphstate\_indep\_qiskit\_3} & Graph State, 3 qubit \\ 
    \texttt{graphstate\_indep\_qiskit\_4} & Graph State, 4 qubit \\ 
    \texttt{graphstate\_indep\_qiskit\_5} & Graph State, 5 qubit \\ 
    \texttt{graphstate\_indep\_qiskit\_6} & Graph State, 6 qubit \\ 
\hline
    \texttt{grover-noancilla\_indep\_qiskit\_2} & Grover's Algorithm (no ancilla), 2 qubit \\    
    \texttt{grover-noancilla\_indep\_qiskit\_3} & Grover's Algorithm (no ancilla), 3 qubit \\    
    \texttt{grover-noancilla\_indep\_qiskit\_4} & Grover's Algorithm (no ancilla), 4 qubit \\    
    \texttt{grover-noancilla\_indep\_qiskit\_5} & Grover's Algorithm (no ancilla), 5 qubit \\    
    \texttt{grover-noancilla\_indep\_qiskit\_6} & Grover's Algorithm (no ancilla), 6 qubit \\
\hline
    \texttt{grover-v-chain\_indep\_qiskit\_2} & Grover's Algorithm (v-chain), 2 qubit \\    
    \texttt{grover-v-chain\_indep\_qiskit\_3} & Grover's Algorithm (v-chain), 3 qubit \\    
    \texttt{grover-v-chain\_indep\_qiskit\_4} & Grover's Algorithm (v-chain), 4 qubit \\    
    \texttt{grover-v-chain\_indep\_qiskit\_5} & Grover's Algorithm (v-chain), 5 qubit \\
\hline
    \texttt{portfolioqaoa\_indep\_qiskit\_3} & Portfolio Optimization with QAOA, 3 qubit  \\    
    \texttt{portfolioqaoa\_indep\_qiskit\_4} & Portfolio Optimization with QAOA, 4 qubit  \\    
    \texttt{portfolioqaoa\_indep\_qiskit\_5} & Portfolio Optimization with QAOA, 5 qubit  \\    
    \texttt{portfolioqaoa\_indep\_qiskit\_6} & Portfolio Optimization with QAOA, 6 qubit  \\
\hline
    \texttt{portfoliovqe\_indep\_qiskit\_3} & Portfolio Optimization with VQE, 3 qubit  \\    
    \texttt{portfoliovqe\_indep\_qiskit\_4} & Portfolio Optimization with VQE, 4 qubit  \\    
    \texttt{portfoliovqe\_indep\_qiskit\_5} & Portfolio Optimization with VQE, 5 qubit  \\    
    \texttt{portfoliovqe\_indep\_qiskit\_6} & Portfolio Optimization with VQE, 6 qubit  \\
\hline
    \texttt{qaoa\_indep\_qiskit\_3} & Quantum Approximation Optimization Algorithm (QAOA), 3 qubit  \\     
    \texttt{qaoa\_indep\_qiskit\_4} & Quantum Approximation Optimization Algorithm (QAOA), 4 qubit  \\    
    \texttt{qaoa\_indep\_qiskit\_5} & Quantum Approximation Optimization Algorithm (QAOA), 5 qubit  \\    
    \texttt{qaoa\_indep\_qiskit\_6} & Quantum Approximation Optimization Algorithm (QAOA), 6 qubit  \\
\hline
\hline
\end{tabular}  
\end{table}

\begin{table}[thb]
\small
\renewcommand{\arraystretch}{1.0}
\begin{tabular}{||l|l||}
\hline
\hline
Benchmark Codename & Circuit Name and Size \\
\hline
\hline
    \texttt{qft\_indep\_qiskit\_2} & Quantum Fourier Transformation (QFT), 2 qubit \\    
    \texttt{qft\_indep\_qiskit\_3} & Quantum Fourier Transformation (QFT), 3 qubit \\    
    \texttt{qft\_indep\_qiskit\_4} & Quantum Fourier Transformation (QFT), 4 qubit \\    
    \texttt{qft\_indep\_qiskit\_5} & Quantum Fourier Transformation (QFT), 5 qubit \\    
    \texttt{qft\_indep\_qiskit\_6} & Quantum Fourier Transformation (QFT), 6 qubit \\
\hline
    \texttt{qftentangled\_indep\_qiskit\_2} & Entangled QFT, 2 qubit \\    
    \texttt{qftentangled\_indep\_qiskit\_3} & Entangled QFT, 3 qubit \\    
    \texttt{qftentangled\_indep\_qiskit\_4} & Entangled QFT, 4 qubit \\    
    \texttt{qftentangled\_indep\_qiskit\_5} & Entangled QFT, 5 qubit \\    
    \texttt{qftentangled\_indep\_qiskit\_6} & Entangled QFT, 6 qubit \\ 
\hline
    \texttt{qnn\_indep\_qiskit\_2} & Quantum Neural Network (QNN), 2 qubit \\    
    \texttt{qnn\_indep\_qiskit\_3} & Quantum Neural Network (QNN), 3 qubit \\    
    \texttt{qnn\_indep\_qiskit\_4} & Quantum Neural Network (QNN), 4 qubit \\    
    \texttt{qnn\_indep\_qiskit\_5} & Quantum Neural Network (QNN), 5 qubit \\    
    \texttt{qnn\_indep\_qiskit\_6} & Quantum Neural Network (QNN), 6 qubit \\
\hline
    \texttt{qpeexact\_indep\_qiskit\_2} & Quantum Phase Estimation (QPE) exact, 2 qubit \\    
    \texttt{qpeexact\_indep\_qiskit\_3} & Quantum Phase Estimation (QPE) exact, 3 qubit \\    
    \texttt{qpeexact\_indep\_qiskit\_4} & Quantum Phase Estimation (QPE) exact, 4 qubit \\    
    \texttt{qpeexact\_indep\_qiskit\_5} & Quantum Phase Estimation (QPE) exact, 5 qubit \\    
    \texttt{qpeexact\_indep\_qiskit\_6} & Quantum Phase Estimation (QPE) exact, 6 qubit \\
\hline
    \texttt{qpeinexact\_indep\_qiskit\_2} & Quantum Phase Estimation (QPE) inexact, 2 qubit \\    
    \texttt{qpeinexact\_indep\_qiskit\_3} & Quantum Phase Estimation (QPE) inexact, 3 qubit \\    
    \texttt{qpeinexact\_indep\_qiskit\_4} & Quantum Phase Estimation (QPE) inexact, 4 qubit \\    
    \texttt{qpeinexact\_indep\_qiskit\_5} & Quantum Phase Estimation (QPE) inexact, 5 qubit \\    
    \texttt{qpeinexact\_indep\_qiskit\_6} & Quantum Phase Estimation (QPE) inexact, 6 qubit \\
\hline
    \texttt{qwalk-noancilla\_indep\_qiskit\_3} & Quantum Walk (no ancilla), 3 qubit \\    
    \texttt{qwalk-noancilla\_indep\_qiskit\_4} & Quantum Walk (no ancilla), 4 qubit \\    
    \texttt{qwalk-noancilla\_indep\_qiskit\_5} & Quantum Walk (no ancilla), 5 qubit \\    
    \texttt{qwalk-noancilla\_indep\_qiskit\_6} & Quantum Walk (no ancilla), 6 qubit \\
\hline
    \texttt{qwalk-v-chain\_indep\_qiskit\_3} & Quantum Walk (v-chain), 3 qubit \\    
    \texttt{qwalk-v-chain\_indep\_qiskit\_5} & Quantum Walk (v-chain), 5 qubit \\
\hline
    \texttt{vqe\_indep\_qiskit\_3} & Variational Quantum Eigensolver (VQE), 3 qubit \\    
    \texttt{vqe\_indep\_qiskit\_5} & Variational Quantum Eigensolver (VQE), 5 qubit \\    
    \texttt{vqe\_indep\_qiskit\_6} & Variational Quantum Eigensolver (VQE), 6 qubit \\
\hline
    \texttt{wstate\_indep\_qiskit\_2} & W-State, 2 qubit \\    
    \texttt{wstate\_indep\_qiskit\_3} & W-State, 3 qubit \\    
    \texttt{wstate\_indep\_qiskit\_4} & W-State, 4 qubit \\    
    \texttt{wstate\_indep\_qiskit\_5} & W-State, 5 qubit \\    
    \texttt{wstate\_indep\_qiskit\_6} & W-State, 6 qubit \\
\hline
\hline
\end{tabular}  
\end{table}


\end{document}